\documentstyle[aps,epsfig,multicol]{revtex}
\begin{document}

\def\breakon{\end{multicols}\widetext\vspace{-.2cm}
\noindent\rule{.48\linewidth}{.3mm}\rule{.3mm}{.3cm}\vspace{.0cm}}

\def\breakoff{\vspace{-.2cm}
\noindent
\rule{.52\linewidth}{.0mm}\rule[-.27cm]{.3mm}{.3cm}\rule{.48\linewidth}{.3mm}
\vspace{-.3cm}
\begin{multicols}{2}
\narrowtext}

\title{Superconductors with Magnetic Impurities: Instantons and Sub-gap States}

\author{A Lamacraft and B D Simons} 
\address{Cavendish Laboratory, Madingley Road, Cambridge CB3\ OHE, UK}

\date{\today}

\maketitle 

\begin{abstract}
When subject to a weak magnetic impurity potential, the order parameter
and quasi-particle energy gap of a bulk singlet superconductor are suppressed.
According to the conventional mean-field theory of Abrikosov and Gor'kov,
the integrity of the energy gap is maintained up to a critical concentration
of magnetic impurities. In this paper, a field theoretic approach is 
developed to critically analyze the validity of the mean field theory. 
Using the supersymmetry technique we find a spatially homogeneous 
saddle-point that reproduces the Abrikosov-Gor'kov theory, and identify 
instanton contributions to the density of states that render the
quasi-particle energy gap soft at any non-zero magnetic impurity
concentration. The sub-gap states are associated with supersymmetry broken
field configurations of the action. An analysis of fluctuations around
these configurations shows how the underlying supersymmetry of the action
is restored by zero modes. An estimate of the density of states is given
for all dimensionalities. To illustrate the universality of the
present scheme we apply the same method to study `gap
fluctuations' in a normal quantum dot coupled to a superconducting
terminal. Using the same instanton approach, we recover the universal
result recently proposed by Vavilov \emph{et al}. Finally, we emphasize
the universality of the present scheme for the description of gap 
fluctuations in $d$-dimensional superconducting/normal structures.
\end{abstract}

\bigskip




\begin{multicols}{2}
\narrowtext

\section{Introduction}

While the spectral properties of a singlet $s$-wave superconductor are largely 
unaffected by weak non-magnetic impurities~\cite{anderson}, the pair-breaking
effect of magnetic impurities leads to the gradual destruction of 
superconductivity. Remarkably, the suppression of the quasi-particle energy 
gap is more rapid than that of the superconducting order parameter, 
admitting the existence of a narrow `gapless' superconducting phase~\cite{ag}
in which the quasi-particle energy gap is destroyed while the superconducting
order parameter remains non-zero. Now, according to the conventional 
(mean-field) description formulated in the seminal work of Abrikosov and 
Gor'kov (AG), an energy gap is maintained up to a critical concentration 
of magnetic impurities (at $T=0$, $91$\% of the critical concentration at
which superconductivity is destroyed). Yet, being unprotected by the 
Anderson theorem, it seems likely that the gap structure predicted by the 
mean-field theory is untenable and must be destroyed by `optimal' 
fluctuations of the random impurity potential. Indeed, since the pioneering 
work of AG, several authors~\cite{yl,shiba,rus,makires,bt,bna} have explored 
the nature of `sub-gap' states in the superconducting system. The aim of 
this work is to present a detailed investigation of the spectrum and 
profile of sub-gap states in superconductors subject to a weak magnetic 
and non-magnetic impurity potential, thus systematically improving
upon the mean-field theory of AG. Our preliminary findings have
already been reported in a recent letter~\cite{lslett}.

In the earliest works on the subject~\cite{yl,shiba,rus}, attention was 
focussed on the the influence of strong magnetic impurities. In particular, 
in the unitarity limit, it was shown that a single magnetic impurity leads 
to the local suppression of the order parameter and creates a bound sub-gap 
quasi-particle state~\cite{yl}.
%
%
For a finite impurity concentration, these intra-gap states broaden into 
a band~\cite{shiba} merging smoothly with the continuum bulk states. 

By contrast, starting with a {\em weak} magnetic impurity distribution 
(i.e. one in which the magnetic scattering can be treated within the 
Born approximation), the mean-field theory of AG~\cite{ag} predicts a 
gradual suppression of the quasi-particle energy gap. Defining the 
dimensionless parameter
\begin{eqnarray}
\zeta\equiv {1\over\tau_s|\Delta|}, 
\label{eq:zeta}
\end{eqnarray}
where $|\Delta|$ represents the value of the homogeneous self-consistent 
order parameter, and $\tau_s$ denotes the Born scattering time due to 
magnetic impurities, the AG theory shows the gap to follow the relation
\begin{eqnarray}
E_{\rm gap}(\tau_s)=|\Delta|\left(1-\zeta^{2/3}\right)^{3/2}
\label{eq:egap}
\end{eqnarray}
showing an onset of the gapless region at $\zeta=1$ (note $\hbar=1$ 
throughout). Within the same mean-field theory, for $\zeta\le 1$, the 
self-consistent order parameter varies as $|\Delta|=|\bar{\Delta}|
\exp[-\pi\zeta/4]$ where $|\bar{\Delta}|$ represents the order parameter
of the clean superconductor, confirming that the order parameter is finite
at the onset of the gapless phase. The precise variation of the 
quasi-particle energy gap is compared to that of the self-consistent order 
parameter in Fig.~\ref{fig:gap_delta}. Staying within the framework of
the mean-field theory, one can obtain a smooth interpolation between the 
strong and weak impurity scattering behaviors~\cite{shiba}.

\begin{figure}[hbtp]
\centerline{\epsfxsize=3in\epsfbox{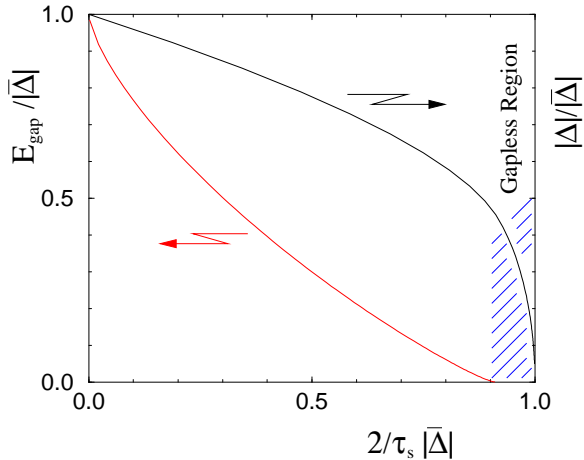}}\bigskip
\caption{Variation of the energy gap $E_{\rm gap}$ and the self-consistent
order parameter $|\Delta|$ as a function of (normalized) scattering
rate $2/\tau_{\rm s}|\bar{\Delta}|$.}
\label{fig:gap_delta}
\end{figure}

However, even for weak disorder, it is apparent that optimal 
fluctuations of the random potential must generate sub-gap states in the 
interval $0<\zeta<1$, and therefore provide non-perturbative corrections
to the self-consistent Born approximation used by AG. Extending the arguments 
of Balatsky and Trugman~\cite{bt}, a fluctuation of the random potential which 
leads to an effective Born scattering rate $1/\tau_s^\prime$ in excess of 
$1/\tau_s$ over a range set by the superconducting coherence length, 
\begin{eqnarray}
\xi=\left({D\over |\Delta|}\right)^{1/2}, 
\label{eq:xi}
\end{eqnarray}
induces quasi-particle states down to energies $E_{\rm gap}(\tau_s^\prime)$. 
These sub-gap states are localized, being bound to the region or `droplet' 
where the scattering rate is large. 

The situation bears comparison with band tail states in semi-conductors.
Here rare or optimal configurations of the random impurity potential generate 
bound states, known as Lifshitz tail states~\cite{lifshitz}, which extend
below the band edge. However, the correspondence is, to some extent, 
superficial: band tail states in semi-conductors are typically associated 
with smoothly varying, nodeless wavefunctions. By contrast, the tail states 
below the superconducting gap involve the superposition of states around 
the Fermi level. As such, one expects these states to be rapidly oscillating 
on the scale of the Fermi wavelength $\lambda_F$, but modulated by an 
envelope which is localized on the scale of the coherence length $\xi$. 
This difference is not incidental. Firstly, unlike the semi-conductor, 
one expects the energy dependence of the density of states in the tail 
region below 
the mean-field gap edge to be `universal', independent of the nature of 
the weak impurity distribution, but dependent only on the
pair-breaking parameter $\zeta$. Secondly, as we will see, one can not 
expect a straightforward 
extension of existing theories~\cite{lifshitz,halperin} of the Lifshitz 
tails to describe the profile of tail states in the superconductor.

In the BCS approximation, the random system we consider is specified by the 
Gor'kov Hamiltonian
\begin{equation} 
\hat{H}=\pmatrix{\hat{H}_0 & |\Delta|\sigma^{\sc sp}_2 \cr |\Delta|
\sigma^{\sc sp}_2 & -\hat{H}_0^T}_{\sc ph}
\label{eq:hamiltonian}
\end{equation} 
where the matrix components index the particle/hole content, and
\begin{eqnarray*}
\hat{H}_0={\hat{\bf p}^2\over 2m}-\epsilon_F+V({\bf r})+J{\bf S}({\bf r})\cdot
{\bf \sigma}^{\sc sp}
\end{eqnarray*}
denotes the normal component of the random Hamiltonian. Here we take 
$\Delta=g_\Delta\langle c_\downarrow c_\uparrow\rangle$ to be
spatially homogeneous 
and determined self-consistently from the conventional AG 
mean-field theory. We will assess the validity of this assumption below. 
In addition to the weak potential impurity distribution 
$V({\bf r})$, the particles experience a quenched random magnetic impurity 
distribution $J{\bf S}({\bf r})$ where $J$ represents the exchange coupling 
and Pauli matrices $\{\sigma_i^{\sc sp}\}$ operate on the spin indices. 
The magnetic ${\bf S}({\bf r})$ and non-magnetic $V({\bf r})$ random impurity 
potentials are both taken to be Gaussian $\delta$-correlated with zero 
mean and variance 
\begin{eqnarray*}
\left\langle JS_\alpha({\bf r}) JS_\beta({\bf r}^\prime)
\right\rangle_S &=& {1\over 6\pi\nu\tau_s}\delta^d({\bf r}-{\bf r}^\prime)
\delta_{\alpha\beta}\\
\left\langle V({\bf r}) V({\bf r}^\prime)\right\rangle_V &=&{1\over 2\pi\nu
\tau}\delta^d({\bf r}-{\bf r}^\prime)
\end{eqnarray*}
respectively, and $\nu$ represents the average density of states (DoS) per spin of
the normal system.

To keep our discussion simple, and to make contact with the AG theory, we 
will take the quenched distribution of magnetic impurities to be `classical' 
and non-interacting throughout. For practical purposes, this entails the 
consideration of structures where both the Kondo temperature~\cite{makires} 
and, more significantly, the RKKY induced spin glass temperature~\cite{lmk} 
are smaller than the relevant energy scales of the superconductor. The 
remaining energy scales are arranged in the quasi-classical and dirty 
limits: 
\begin{eqnarray}
\epsilon_F\gg 1/\tau\gg (|\Delta|,1/\tau_s) 
\label{limits}
\end{eqnarray}
where $\tau$ represents the transport time associated with non-magnetic 
impurities. We remark that these limits are not compatible with the situation
in which the magnetic impurities provide the only source of scattering,
i.e. $\xi\sim \ell=v_F\tau_s$. We should not, therefore, expect a 
straightforward comparison with the analysis of Ref.~\cite{bt}.

Within the approximations above we will derive a quasi-classical field theory
of the disordered superconducting system. Following the approach of 
Ref.~\cite{ast} we will express spectral properties of the system in terms
of an intermediate energy scale action which accommodates the quantum 
interference properties of the superconducting system. By investigating 
stationary inhomogeneous instanton field configurations of the action, we 
will expose the structure and profile of the sub-gap states, and thereby
obtain an analytical expression for the DoS to exponential accuracy. 
Finally, we will comment on the universality of the present scheme by 
applying the same technique to investigate `gap fluctuations' in a quantum
dot coupled by open channels to a superconducting terminal. 

Although the analysis is straightforward, the technology is somewhat involved.
We have therefore decided to summarize the main conclusions of this 
investigation here in the introduction. In particular, we will show that,
in the vicinity of the mean-field gap edge, the sub-gap DoS of the 
$d$-dimensional system is dominated by tail states which are confined to 
droplets of size
\begin{eqnarray*}
r_0(\epsilon)\sim {\xi\over (1-\zeta^{2/3})^{1/4}}
\left({|\Delta|\over E_{\rm gap}-\epsilon}\right)^{1/4}.
\end{eqnarray*}
The corresponding sub-gap DoS takes the form
\breakon
\begin{eqnarray*}
{\nu(\epsilon<E_{\rm gap})\over\nu}
\sim\exp\left[-4\pi g (\xi/L)^{d-2} f_d(\zeta) 
\left({E_{\rm gap}-\epsilon\over |\Delta|}
\right)^{(6-d)/4}\right]
\end{eqnarray*}
\breakoff
\noindent
where $g=\nu D L^{d-2}$ denotes the bare dimensionless conductance and 
$f_d(\zeta)=a_d\zeta^{-2/3}(1-\zeta^{2/3})^{-(2+d)/8}$ represents a 
dimensionless function of the control parameter $\zeta$ ($a_d$ const.). 
When reparameterized in terms of the DoS just above the mean-field
gap edge~\cite{Park}
\begin{eqnarray}
\nu(\epsilon>E_{\rm gap})\simeq {1\over \pi L^d}\sqrt{\epsilon-E_{\rm gap}
\over \Delta_g^3}
\label{nuexp}
\end{eqnarray}
where 
\begin{eqnarray*}
\Delta_g^{-3/2}=4\pi\nu L^d \sqrt{2\over 3|\Delta|}\zeta^{-2/3}(1-
\zeta^{2/3})^{-1/4}
\end{eqnarray*}
the expression for the sub-gap DoS can be brought to the more compact form
\begin{eqnarray*}
{\nu(\epsilon<E_{\rm gap})\over\nu}\sim\exp\left[-\widetilde{a}_d
\left({r_0\over L}\right)^d\left({E_{\rm gap}-\epsilon\over 
\Delta_g}\right)^{3/2}\right]
\end{eqnarray*}
with $\widetilde{a}_d$ some numerical constant. 

In the zero-dimensional system, although the interpretation of the optimal 
fluctuation as a localized droplet is no longer appropriate, the expression 
above correctly interpolates to $d=0$ and coincides with the universal 
expression for gap fluctuations proposed in Ref.~\cite{vbab}. The surprising 
dependence of the result on the dimensionless distance from the mean-field 
gap is one of the reasons why a Lifshitz argument appears difficult to 
construct for this problem. 

The paper is organized as follows: in section~\ref{sec:ft} a theory of 
the statistical properties of the Gor'kov Green function is developed
within the framework of a supersymmetric field theory involving
a non-linear $\sigma$-model functional. Here we follow closely the 
analysis of Ref.~\cite{ast} and \cite{bcsz} (c.f. Ref.~\cite{Oppermann}). As a 
result we identify the conventional AG mean-field equation with the 
homogeneous saddle-point 
equation of the effective action. In section~\ref{sec:inhomo} we 
show that the non-vanishing of the DoS beneath the gap 
predicted by the standard AG theory is due to the appearance of 
inhomogeneous ``instanton'' saddle points of finite action. We identify 
the profile of these instantons with the envelope modulating the 
quasi-classical sub-gap states. The instanton configurations considered 
break the underlying supersymmetry of the action, and a detailed 
examination of fluctuations is required to understand how supersymmetry 
is restored by zero modes, as well as to appreciate fully how these 
configurations are able to contribute to the DoS. In 
section~\ref{sec:zerod} we examine the zero dimensional limit of the 
problem and compare our results to the recent literature~\cite{bna,vbab}.
In doing so, we provide an explanation of the universal results 
reported in Ref.~\cite{vbab}, and discuss the universality of the
$d>0$ result. Finally,
in section~\ref{sec:discuss} we speculate on potential generalizations
of the results presented here.

\section{Field Theory of the Disordered Superconductor}
\label{sec:ft}

The construction of the field theory of the disordered superconductor 
follows the quasi-classical method of Eilenberger~\cite{eilenberger} 
and Usadel~\cite{usadel} elevated to the level of an effective action. 
The starting point of the analysis is the generating functional for 
the single-particle Gor'kov Green function for the non-interacting 
quasi-particle Bogoluibov Hamiltonian. Here we borrow our notation
from Ref.~\cite{bcsz}.

\subsection{Generating Functional}

Single-particle properties of the Gor'kov Hamiltonian~(\ref{eq:hamiltonian}) 
are obtained from the generating functional 
\begin{equation} 
{\cal Z}[J] = \int D(\bar\psi,\psi) e^{\int d{\bf r}\;\left(i\bar\psi 
(\hat{H}-\epsilon_-)\psi+\bar\psi J +\bar J \psi\right)},
\label{eq:genfn}
\end{equation}
where $\epsilon_-\equiv\epsilon-i0$ and the supervector fields have the 
internal structure ${\bar{\psi}=\left(
\matrix{\bar{\psi}_\uparrow & \bar{\psi}_\downarrow & 
\psi_\uparrow & \psi_\downarrow}\right)}$, $\psi^T=\left(\matrix{\psi_\uparrow 
& \psi_\downarrow & \bar{\psi}_\uparrow &  \bar{\psi}_\downarrow}\right)$.
As usual, by introducing both commuting and anticommuting elements the 
normalization ${\cal Z}[0]=1$ is assured. (The generalization to the 
consideration of higher point functions, which involves the extension of 
the field space, follows straightforwardly.)

To condense the notation, it is convenient to perform the rotation
$\psi\mapsto \psi^\prime=U\psi$, $\bar{\psi}\mapsto \bar{\psi}^\prime=
\bar{\psi}U^\dagger$ with
\begin{eqnarray*}
U=\pmatrix{ 1 & 0 \cr 0 & i\sigma^{\sc sp}_2 }_{\sc ph} \;,
\end{eqnarray*}
after which the Gor'kov Hamiltonian takes the form
\begin{eqnarray*}
\hat{H}=\left({\hat{\bf p}^2\over 2m}+V({\bf r})-\epsilon_F\right)\otimes
\sigma^{\sc ph}_3+J{\bf S}({\bf r})\cdot {\bf \sigma^{\sc sp}}+|\Delta|
\sigma^{\sc ph}_2 \;.
\end{eqnarray*}

The unusual phase coherence properties of the superconducting system rely
on the particle/hole or charge conjugation symmetry
\begin{equation} 
\hat{H}=-\sigma^{\sc ph}_2\otimes\sigma^{\sc sp}_2 \hat{H}^T\sigma^{\sc sp}_2
\otimes\sigma^{\sc ph}_2.
\label{eq:newphsym}
\end{equation}
To easily accommodate these effects, it is convenient to introduce the further 
space doubling,
\begin{eqnarray} 
&&2\bar \psi (\hat{H}-\epsilon_-)\psi=\bar \psi (\hat{H}-\epsilon_-)\psi
 + \psi^T (\hat{H}^T-\epsilon_-) \bar \psi^T\nonumber\\
&&=\bar \psi (\hat{H}-\epsilon_-)\psi + \psi^T(
\sigma^{\sc ph}_2\otimes\sigma^{\sc sp}_2 \hat{H}\sigma^{\sc sp}_2\otimes
\sigma^{\sc ph}_2-\epsilon_-)\bar \psi ^T\nonumber\\
&&=2\bar \Psi (\hat{H}-\epsilon_-\sigma^{\sc cc}_3)\Psi
\label{eq:softdouble}
\end{eqnarray}
where, defining the Pauli matrix $\sigma_3^{\sc cc}$ which operates in the 
charge-conjugation ({\sc cc}) space,
\begin{eqnarray*}
&&\Psi=\frac{1}{\sqrt{2}}
\pmatrix{\psi\cr \sigma^{\sc ph}_2\otimes\sigma^{\sc sp}_2\bar \psi ^T
\cr}_{\sc cc},\\
&&\bar\Psi = \frac{1}{\sqrt{2}}
\pmatrix{\bar\psi & -\psi^T \sigma^{\sc ph}_2\otimes
\sigma^{\sc sp}_2}_{\sc cc}.
\end{eqnarray*}
This completes the formulation of the generating functional
as a field integral involving $16$-component supervector fields $\Psi$ and 
$\bar \Psi$. The latter obey the symmetry relations
\begin{eqnarray} 
\Psi = -\sigma^{\sc ph}_2\otimes\sigma^{\sc sp}_2 \gamma \,\bar \Psi^T,\qquad 
\bar\Psi =  \Psi^T \sigma^{\sc ph}_2\otimes\sigma^{\sc sp}_2 \gamma^{-1}
\label{eq:psisym}
\end{eqnarray}
with $\gamma = E_{\sc bb} i\sigma^{\sc cc}_2-E_{\sc ff} \sigma^{\sc cc}_1$, 
where $E_{\sc bb}={\rm diag}(1,0)_{\sc bf}$ and 
$E_{\sc ff}={\rm diag}(0,1)_{\sc bf}$ project into the Boson-Boson and 
Fermion-Fermion sectors respectively.

\subsection{Ensemble Averaging}

Cast as a field integral, the impurity average of the generating functional
over the Gaussian distributed random impurity potentials is straightforward.
Separating the regular from the disordered components of the Hamiltonian 
$\hat{H}=\hat{\cal H}_0+V({\bf r})\sigma^{\sc ph}_3+J{\bf S}({\bf r})\cdot 
{\bf \sigma^{\sc sp}}$, an ensemble average over the random potentials
obtains
\breakon
\begin{eqnarray} 
\left\langle{\cal Z}[0]\right\rangle_{V,S}=\int D(\bar \Psi,\Psi) \exp\left[
\int d{\bf r} \left( i \bar \Psi (\hat{\cal H}_0-\epsilon_-\sigma^{\sc cc}_3)
\Psi-\frac{1}{4\pi\nu\tau}( \bar \Psi \sigma^{\sc ph}_3 \Psi)^2 -
\frac{1}{12\pi\nu\tau_s}( \bar \Psi {\bf \sigma^{\sc sp}} \Psi)^2\right)
\right].
\label{eq:avgenfn}
\end{eqnarray}
\breakoff

The interactions generating by the impurity averaging can be decoupled by the
introduction of a Hubbard-Stratonovich field. Beginning with the non-magnetic
disorder, slow modes of the action are identified by rewriting the action in
the approximate form
\begin{eqnarray*}
\frac{1}{4\pi\nu\tau} \int d{\bf r}\;\left(\bar \Psi \sigma^{\sc ph}_3 
\Psi\right)^2 \simeq \frac{1}{2\pi\nu\tau} \sum_{|{\bf q}| < \ell^{-1}} 
{\rm str}\; \zeta(-{\bf q}) \zeta({\bf q}),
\end{eqnarray*}
where $\zeta({\bf q})=\sum_{\bf k} \Psi({\bf k})\otimes\bar \Psi(-{\bf k}+
{\bf q})\sigma^{\sc ph}_3$. The latter can be decoupled by the slowly
varying $16\times 16$ supermatrix field $Q({\bf r})$ according to the 
identity
\begin{eqnarray*}
&&\exp\left[-\frac{1}{2\pi\nu\tau}\sum_{\bf q} {\rm str}\; \zeta({\bf q}) 
\zeta(-{\bf q})\right]\\ 
&&=\int DQ \exp\left[\sum_{\bf q} {\rm str}\left(\frac{\pi \nu}
{8\tau} Q({\bf q})Q(-{\bf q})-{1\over 2\tau}Q({\bf q})\zeta(-{\bf q})
\right)\right].
\end{eqnarray*}
The symmetry properties of the fields $Q({\bf r})$ are inherited from
the dyadic product $\Psi({\bf r})\otimes\bar\Psi({\bf r})\sigma_3^{\sc ph}$.
Making use of Eq.~(\ref{eq:psisym}) one finds the symmetry relations
\begin{equation} 
Q=\sigma^{\sc ph}_1\otimes\sigma^{\sc sp}_2\gamma Q^T \gamma^{-1} 
\sigma^{\sc ph}_1\otimes\sigma^{\sc sp}_2.
\label{eq:Qsym}
\end{equation}

The interaction generated by the magnetic impurity averaging can be 
treated~\cite{efetov} by performing all possible pairings and making use
of the saddle-point approximation $Q({\bf r})=2\langle \Psi({\bf r})
\otimes\bar\Psi({\bf r})\sigma^{\sc ph}_3\rangle_\Psi/\pi\nu$. This leads 
to the replacement 
\begin{eqnarray*}
&&\frac{1}{12\pi\nu\tau_s}\int d{\bf r}\; \left(\bar \Psi 
{\bf \sigma^{\sc sp}} \Psi\right)^2\\
&&\qquad\qquad\qquad\mapsto \frac{\pi\nu}{24\tau_s}
\int d{\bf r}\; {\rm str}\;\left(
Q \sigma^{\sc ph}_3\otimes {\bf \sigma^{\sc sp}}\right)^2.
\end{eqnarray*}
Such an approximation, which neglects pairings at non-coincident points is
allowed by the strong inequality $(\ell/\xi)^d\ll 1$. In addition we 
discard the contraction $\langle\bar\Psi{\bf \sigma^{\sc sp}}\Psi
\rangle_\Psi$. The term generated by this procedure could in any case
be decoupled by a slow Bosonic field ${\bf S}({\bf r})$ which would 
immediately be set to zero for the singlet saddle-points that will be the 
basis of this paper.

Gaussian in the fields $\Psi$ and $\bar\Psi$, the functional integration can
be performed explicitly after which one obtains $\langle{\cal Z}[0]
\rangle_{V,S}=\int DQ \exp(-S[Q])$ where 
\breakon
\begin{eqnarray*}
S[Q]=-\int d{\bf r}\left[\frac{\pi \nu}{8 \tau}{\rm str}\; Q^2 -
\frac{1}{2}{\rm str}\;\ln \left(\sigma^{\sc ph}_3(\hat{\cal H}_0-
\epsilon_-\sigma^{\sc cc}_3)+ 
{i\over 2\tau} Q\right) -\frac{\pi\nu}{24\tau_s} 
{\rm str}\;(Q \sigma^{\sc ph}_3\otimes {\bf \sigma^{\sc sp}})^2 \right].
\end{eqnarray*}
\breakoff
\noindent
Further progress is possible only within a saddle-point approximation.
Following Ref.~\cite{ast}, the saddle-point analysis will be conducted
in a two-step process:

\subsection{Saddle-point Approximation and the Non-linear $\sigma$-model}

The first task is to make use of the quasi-classical parameter $\epsilon_F\tau
\gg 1$ to construct an intermediate energy scale action. Dropping the 
symmetry breaking perturbations $\epsilon$, $|\Delta|$ and $1/\tau_s$, a 
variation of the action with respect to fluctuations of $Q$ obtains the 
saddle-point equation
\begin{eqnarray*}
Q({\bf r})={i\over \pi\nu}\langle{\bf r}|[i0\sigma_3^{\sc cc}\otimes
\sigma_3^{\sc ph}+\hat{h}_0+iQ/2\tau]^{-1}|{\bf r}\rangle\;,
\end{eqnarray*}
where $\hat{h}_0=\hat{\bf{p}}^2/2m-\epsilon_{\sc f}$. Taking into account 
the analytical properties of the Green function and the
quasi-classical limit $\epsilon_{\sc f}\tau\gg 1$, one
obtains the solution $Q_0=\sigma_3^{\sc ph}\otimes\sigma_3^{\sc
  cc}$. However, as usual, this saddle-point is not unique. In
particular, $Q=TQ_0T^{-1}$ is also a solution, for any constant matrix
$T$ which is consistent with the fundamental symmetry (\ref{eq:Qsym})
for $Q$. Transverse fluctuations of $Q$ away from the
$Q^2=\openone$ manifold may be integrated out within the saddle-point 
approximation due to the large parameter $\nu L^d/\tau\gg 1$, where
$L$ is the system size~\cite{efetov}. Restricting attention to the
manifold generated by the non-linear constraint $Q^2=\openone$, an
effective action is obtained by allowing $T$ to vary in space and
expanding to second order in gradients of $T$, and first order in 
$\epsilon$, $|\Delta|$, and $1/\tau_s$
\breakon
\begin{equation} 
S[Q] = -\frac{\pi \nu}{8} \int d{\bf r}\; {\rm str}\; \left[D(\partial
  Q)^2 -4i 
\left(\epsilon_-\sigma^{\sc cc}_3+|\Delta|\sigma^{\sc ph}_2\right)
\sigma^{\sc ph}_3 Q-\frac{1}{3\tau_s} \left(Q\sigma^{\sc ph}_3\otimes 
{\bf \sigma^{\sc sp}}\right)^2\right].
\label{eq:sigmod}
\end{equation}
\breakoff
\noindent
Here $D=v_F^2\tau/d$ represents the classical diffusion constant associated 
with the non-magnetic impurities. In particular, the quasi-particle DoS
is obtained from the functional integral
\begin{eqnarray}
\langle\nu(\epsilon,{\bf r})\rangle_{V,S}={\nu\over 4}{\rm Re}\left\langle 
{\rm str}\left(\sigma_3^{\sc bf}\otimes\sigma_3^{\sc ph}\otimes
\sigma_3^{\sc cc}Q({\bf r})\right)\right\rangle_Q.
\label{eq:dos_source}
\end{eqnarray}
The numerical factor leads to a DoS of $4\nu$ for the system as 
$|\epsilon|\to\infty$. This is because both the particle-hole 
structure of the original Bogoliubov Hamiltonian 
and the spin each cause a doubling of the DoS.

This completes the derivation of the intermediate energy scale action.
In the following section we will investigate the action (\ref{eq:sigmod}) 
within mean-field theory and the influence of soft fluctuations on the 
low-energy properties of the system.

\subsection{AG Mean-Field Theory and Fluctuations} \label{sec:ag}

The presence of symmetry-breaking terms in (\ref{eq:sigmod})
originating from the order parameter and the magnetic impurity
potential means that a mean-field analysis is already non-trivial. To 
assimilate the effect of these terms, and to establish contact with
the AG theory, it is necessary to explore the saddle-point equation. 

Varying the non-linear $\sigma$-model action with respect to 
fluctuations of $Q$, subject to the non-linear constraint, one obtains 
the saddle-point or mean-field equation,
\begin{eqnarray*}
&&D \partial\left(Q\partial Q\right)+i\left[Q,\epsilon_-\sigma_3^{\sc cc}
\otimes\sigma_3^{\sc ph}+i|\Delta|\sigma_1^{\sc ph}\right]\nonumber\\
&&\qquad \qquad +\frac{1}{6\tau_s}\left[Q,\sigma_3^{\sc ph}\otimes
{\bf \sigma}^{\sc sp} Q\sigma_3^{\sc ph}\otimes
{\bf \sigma}^{\sc sp}\right] = 0.
\end{eqnarray*}
With the {\em Ansatz}: 
\begin{eqnarray*}
Q_{\sc mf}=\left[\sigma_3^{\sc cc}\otimes\sigma_3^{\sc ph}\cosh\hat\theta+
i\sigma_1^{\sc ph}\sinh\hat\theta\right]\otimes\openone^{\sc sp}, 
\end{eqnarray*}
where the elements $\hat\theta={\rm diag}(\theta_1,i\theta)_{\sc bf}$ are
diagonal in the superspace, the saddle-point equation decouples into 
Boson-Boson and Fermion-Fermion sectors, and takes the form
\begin{eqnarray}
\partial_{{\bf r}/\xi}^2\hat\theta+2i\left(\cosh\hat\theta-\frac{\epsilon}
{|\Delta|}\sinh\hat\theta\right)-\zeta\sinh(2\hat\theta)=0 \;,
\label{eq:usadel}
\end{eqnarray}
a result reminiscent of the Usadel equation of quasi-classical 
superconductivity~\cite{usadel}. This is no coincidence: when 
subject to an inhomogeneous order parameter, the same effective 
action~(\ref{eq:sigmod}) describes the proximity effect in a hybrid 
normal/superconducting compound~\cite{ast}. In the present context,
the spatially homogeneous form of Eq.~(\ref{eq:usadel}) should be 
combined with the self-consistent equation for the order parameter 
\begin{eqnarray*}
|\Delta|=\frac{\pi\nu g_{\Delta}}{\beta}\sum_n \sin\theta_n \;.
\end{eqnarray*}
Here $g_{\Delta}$ is the BCS coupling constant and $\theta_n$ indicates 
that the solution of the Usadel equation is taken at Matsubara frequency 
$\epsilon\to i\epsilon_n$. These two equations then coincide with the 
mean-field equations obtained by AG, and correctly reproduce the known
form of $E_{\rm gap}$ as specified by Eq.~(\ref{eq:egap}).

The AG solution is not unique: for $\epsilon\to 0$, the saddle-point 
equation admits an entire manifold of homogeneous solutions parameterized 
by the transformations $Q=TQ_{\sc mf}T^{-1}$ where now $T=\openone_{\sc ph}
\otimes\openone_{\sc sp}\otimes t$ and $t=\gamma(t^{-1})^T\gamma^{-1}$. 
Soft fluctuations of the fields, which are controlled by a non-linear 
$\sigma$-model defined on the manifold $T\in {\rm OSp}(2|2)/
{\rm GL}(1|1)$ (symmetry class D in the classification of 
Ref.~\cite{zirn_class}), control the low-energy, long-range 
properties of the gapless system giving rise to unusual localization and 
spectral properties. (For a comprehensive discussion of 
the physics of the gapless phase, we refer to 
Refs.~\cite{az,bcsz2,sf,rg,bsz}.) The phenomenology of Class D in the 
context of the gapless phase of the present system will be the subject of a 
forthcoming paper~\cite{ls}. Here we focus on the gapped phase, where 
Class D fluctuations are just one of a host of massive modes that are 
unimportant in the description of sub-gap states.

This completes the formal description of the bulk superconducting
phase. The solution of the AG mean-field equation provides an adequate 
description of the bulk extended states. Soft fluctuations around the AG 
mean-field describe phase coherence effects due to quantum interference. 
However, within the present scheme it is not yet clear 
how to accommodate sub-gap states in the gapped phase of the 
AG theory. To identify such states, it is necessary to return to the 
saddle-point equation~(\ref{eq:usadel}) and seek spatially {\em inhomogeneous} 
solutions.

\section{Instantons and sub-gap states}
\label{sec:inhomo}

Although the reduction and eventual destruction of the quasi-particle 
energy gap predicted by the AG mean-field theory can be reasonably justified 
on purely physical grounds, the integrity of the gap of the range $0<\zeta<1$
is less credible. Once time-reversal is symmetry is broken and the protection 
of Anderson's theorem is lost, there remains no reason why a sharp gap 
should persist. Add to this the observation that the spin scattering rate 
must be subject to spatial fluctuations from the average value 
$1/\tau_s$, and one concludes that corrections to the DoS predicted by the 
AG theory must lead to the appearance of sub-gap states analogous to 
``band tails'' in a disordered semiconductor~\cite{lifshitz,halperin}. 

This analogy is of course not new~\cite{bt,bna} nor, as far as practical 
calculation in the present formulation is concerned, is it particularly 
deep. This is because all averages have already been taken, so we can not 
look for an optimal fluctuation of some potential, as in the classic 
approaches to the study of band tail states in disordered 
semi-conductors~\cite{halperin}.
However, these studies hint at how one can proceed.

Band tail states in semi-conductors can be studied within the same 
functional integral formulation. In particular, the generating function
of the single-particle Green function of a normal disordered conductor
can be presented in the form of a supersymmetric field integral
\begin{eqnarray*}
{\cal Z}[0]=\int D(\Psi,\bar\Psi) \exp\left[i\int d{\bf r}\bar\Psi
\left(\epsilon_+-{\hat{\bf p}^2\over 2m}-V({\bf r})\right)
\Psi\right],
\end{eqnarray*}
where, once again, the random impurity distribution is drawn from 
a Gaussian $\delta$-correlated white-noise impurity potential. The optimal
fluctuation method involves minimizing the action with respect to 
fluctuations in the fields $\Psi$ and potential $V$. This involves
seeking inhomogeneous solutions of the non-linear Schr\"odinger equation
\begin{eqnarray*}
\left(\epsilon-{\hat{\bf p}^2\over 2m}-V({\bf r})\right)\Psi=0,
\end{eqnarray*}
where the corresponding optimal potential is determined self-consistently
by the relation $V({\bf r})=-|\Psi({\bf r})|^2/2\pi\nu\tau$.
In the supersymmetric formulation, band tail states are identified with
supersymmetry broken inhomogeneous solutions of the saddle-point equation 
(see Cardy~\cite{cardy} and Affleck~\cite{affleck}). Indeed, the
anticipated exponential suppression of the DoS necessitates a breaking
of supersymmetry to support a finite action.
Here the phrase ``supersymmetry breaking'' is potentially misleading. 
We use it only to refer to \emph{field configurations}, ubiquitous 
in the problems under discussion here, that do not respect the parity 
between Bose and Fermi degrees of freedom. However, any such 
configuration is just one member of a degenerate manifold differing by 
supersymmetric transformations. The latter maintain the invariance of 
the generating functional ${\cal Z}[0]$ under global supersymmetric 
transformations.

What does this tell us about the identification of optimal fluctuations 
and sub-gap states in the superconductor? Following the analysis above, 
one might guess that sub-gap states are associated with inhomogeneous 
configurations of the $\Psi$ field action. However, we anticipate that 
optimal solutions corresponding to sub-gap states are localized on a 
length scale in excess of the superconducting coherence length. In the 
dirty limit, $\xi\gg \ell\gg \lambda_F$, this implies that the localized 
sub-gap states are quasi-classical in nature. Their 
existence on the level of the $\Psi$ field action will be buried in 
the fast $\lambda_F$ oscillations of the wavefunction. To reveal the
sub-gap states, we must first remove the fast short length scale 
fluctuations of the quasi-classical Green function and look for an 
equation of motion for the slowly varying envelope of the wavefunction.
But this is just the program of the usual quasi-classical method.

The term ``sub-gap states'' is a little misleading in this context. 
Band tails are bound states of some rare potential that sit by themselves 
below the bulk of the spectrum. Each rare configuration that make the
gap soft in the present case will give rise to many states beneath 
the AG gap. Thus the term ``gap fluctuation'', used in
Ref.~\cite{vbab} to describe the zero dimensional SN system, may be 
more appropriate.

As well as being quasi-classical in nature, the existence of sub-gap states 
is not affected by working in the dirty limit. As such, their
existence must be accommodated in the non-linear $\sigma$-model 
functional~(\ref{eq:sigmod}) since the validity of 
this description relied only on the quasi-classical parameter 
$\epsilon_F\tau\gg 1$ and the dirty limit assumption. To identify sub-gap 
states in the present
formalism, we should therefore investigate inhomogeneous solutions 
of the low-energy saddle-point equation in $Q$ --- the Usadel
equation~\cite{usadel}.
Such a solution should be thought of as defining an envelope for the 
quasi-classical sub-gap states.

Therefore, let us revisit the mean-field equation and look for inhomogeneous
solutions at energies $\epsilon<E_{\rm gap}$. To focus our discussion, let
us begin by restricting attention to the quasi one-dimensional geometry. 
To stay firmly within the diffusive regime, we therefore impose the 
requirement that the system size $L$ be much smaller than the localization 
length of the normal system $\xi_{\rm loc.}\simeq \nu L_{\rm w} D$, where 
$L_{\rm w}$ denotes the cross-section. Later, in section~\ref{sec:hidim}, 
we will generalize our discussion to encompass systems of higher dimension.
Furthermore, since, over the interval $0<\zeta<1$, the quasi-particle 
energy gap varies more rapidly than the superconducting order parameter,
we will neglect self-consistency of the order parameter. Taking 
self-consistency into 
account will not alter our qualitative findings, and will only weakly affect 
the quantitative results.

\subsection{Instantons in the Quasi One-dimensional Geometry} 
\label{sec:inst}

To investigate inhomogeneous solutions of the mean-field 
equation~(\ref{eq:usadel}) it is convenient to recast the equation in 
terms of its first integral
\begin{equation} 
(\partial_{x/\xi}\hat\theta)^2+V(\hat{\theta})={\rm const},
\label{eq:firstint}
\end{equation}
where
\begin{eqnarray*}
V(\hat{\theta})=4i\left(\sinh\hat\theta-\frac{\epsilon}{\Delta}
\cosh\hat\theta\right)-\zeta\cosh 2\hat\theta
\end{eqnarray*}
denotes the complex potential. Let us denote by $\theta_{\sc ag}$ the 
values of $\theta_1$ and $i\theta$ at the conventional saddle point, and 
focus on an energy $\epsilon$ below the gap predicted by the AG theory. 
Here ${\rm Im}\;\theta_{\sc ag}=\pi/2$ such that the mean-field DoS 
$\nu_{\sc ag}(\epsilon)=4\nu{\rm Re}\;\cosh\theta_{\sc ag}$ vanishes. The 
corresponding value of ${\rm Re}\;\theta_{\sc ag}$ depends sensitively on 
the energy, with ${\rm Re}\;\theta_{\sc ag}=0$ for $\epsilon=0$. 

Considering the Boson-Boson sector only, if we require that 
$\theta_1(x\to \pm\infty)=\theta_{\rm AG}$, what kind of inhomogeneous
solution is possible? The values of $\theta_1$ at which $\partial_x\theta_1
=0$ can be identified by considering the complex (dimensionless) potential 
function $V(\theta_1)$ from which we can determine the endpoints of the 
`motion' in the complex plane, just as one would use a real potential 
normally. By inspection one may see that, on the line ${\rm Im}\;\theta_1 
= \pi/2$, the potential is purely real. This is not the only contour
where ${\rm Im}\; V=0$, but, by considering forces, it is not hard to see 
that either ${\rm Im}\;\theta_1=\pi/2$ always during the motion, or 
$\theta_1$ follows a trajectory with an endpoint at ${\rm Im}\;\theta_1<0$. 
For reasons outlined below, we will discount this latter possibility. The 
former case amounts to considering ``bounce'' trajectories in the 
\emph{real} potential $V(i\pi/2+\phi)=V_{\sc r}(\phi)$ where
\begin{equation} 
V_{\sc r}(\phi)\equiv -4\left(\cosh\phi-\frac{\epsilon}{|\Delta|}\sinh\phi
\right)+\zeta\cosh 2\phi.
\label{eq:realpot}
\end{equation}
A typical potential is shown in Fig.~\ref{fig:bouncepot}.

\begin{figure}[hbtp]
\centerline{\epsfxsize=3in\epsfbox{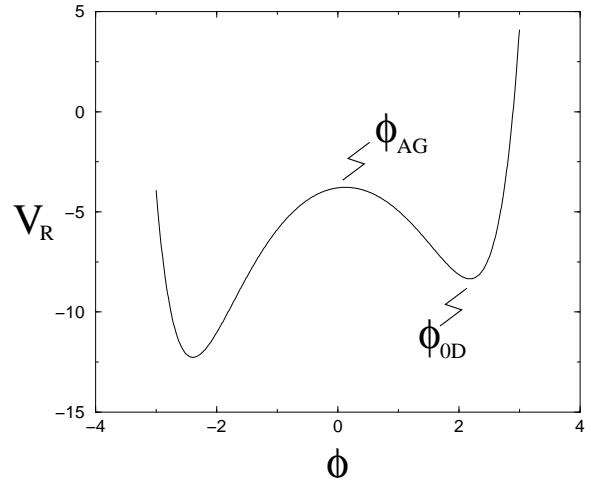}}\bigskip
\caption{Potential $V_{\sc r}(\phi)=V(i\pi/2+\phi)$ for $\epsilon/|\Delta|
=0.1$ and $\zeta=0.2$. The AG saddle point corresponds to the central 
maximum. The saddle point marked $\phi_{\sc 0d}$ is used in the analysis 
of the zero-dimensional problem (section~\ref{sec:zerod}).} 
\label{fig:bouncepot}
\end{figure}

Now integration over the angles $\hat{\theta}$ 
is constrained to certain contours~\cite{efetov}. Is the bounce solution 
accessible to both? As usual, the contour of integration over the Boson-Boson 
field $\theta_1$ includes the entire real axis, while for the Fermion-Fermion 
field, $i\theta$ runs along the imaginary axis from $0$ to $i\pi$. With a 
smooth deformation of the integration contours, the AG saddle-point is 
accessible to both the angles $\hat{\theta}$~\cite{ast}. By contrast, the 
bounce solution {\em and} the AG solution can be reached simultaneously by a 
smooth deformation of the integration contour {\em only} for the Boson-Boson 
field $\theta_1$ (see Fig.~\ref{fig:bounce_contour}). The bounce solution
is therefore associated with a {\em breaking of supersymmetry} at the level 
of the saddle point. 

\begin{figure}[hbtp]
\centerline{\epsfxsize=3in\epsfbox{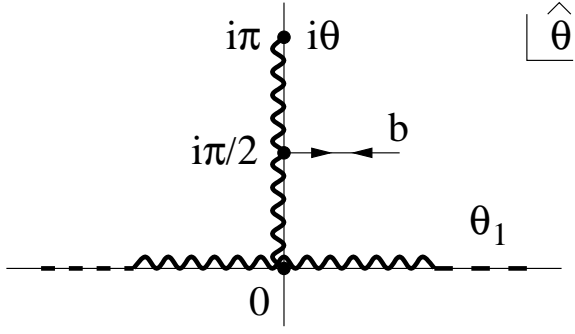}}\bigskip
\caption{Integration contours for Boson-Boson and Fermion-Fermion fields in 
the complex $\hat\theta$ plane. The bounce solution for $\epsilon=0$ 
(labelled as `b') is shown schematically. }
\label{fig:bounce_contour}
\end{figure}

Thus we have identified an inhomogeneous saddle-point configuration for which 
the supersymmetry is broken: $\theta_1$ executes a bounce whilst $i\theta$ 
remains at the mean-field value $\theta_{\sc ag}$. The symmetry broken
solution then incurs the (finite) real action 
\begin{eqnarray*}
S=4\pi\nu L_W (D|\Delta|)^{1/2} S_\phi(\epsilon/|\Delta|, \zeta)
\end{eqnarray*}
where, defining $\phi'$ as the endpoint of the motion,
\begin{equation} 
S_{\phi}\equiv \int_{\phi_{\rm AG}}^{\phi'} 
d\phi \sqrt{V_{\sc r}(\phi_{\sc ag})-V_{\sc r}(\phi)}.
\label{eq:realact}
\end{equation}

Now, as mentioned above, there exists a second possibility for a bounce
solution in which one moves away from $\theta_{\sc ag}$ parallel to the 
imaginary axes. Indeed, such a solution would seem to be a natural 
candidate for the Fermion-Fermion field $i\theta$. However, since the 
endpoint for this trajectory lies at ${\rm Re}\;\theta<0$ outside the 
integration domain which runs from $0$ to $\pi$, this would seem to be 
excluded.

As $\epsilon$ approaches $E_{\rm gap}$ from below, the potential 
(\ref{eq:realpot}) becomes more shallow, with the maximum merging with one 
of the minima when we reach the gap. Near the edge, up to an irrelevant
constant, an expansion of the potential in powers of $(\phi-
\phi_{\sc ag})$ leads to the cubic form
\begin{eqnarray} 
&&V_{\sc r}[\phi]\simeq -\alpha\left(\frac{E_{\rm gap}-\epsilon}{|\Delta|}
\right)^{1/2} (\phi-\phi_{\sc ag})^2\nonumber \\ 
&&\qquad \qquad \qquad \qquad \qquad \qquad + \beta (\phi-\phi_{\sc ag})^3
\label{eq:approxV}
\end{eqnarray}
where the dimensionless coefficients are specified by
\begin{eqnarray} 
\label{eq:coeff}
\alpha=6\sqrt{2\over 3}\left(\frac{E_{\rm gap}}{|\Delta|}\right)^{1/6}, 
\qquad 
\beta=2\left(\frac{\zeta E_{\rm gap}}{|\Delta|}\right)^{1/3}.
\end{eqnarray}
Note that, making use of Eq.~(\ref{eq:egap}), both of these
coefficients depend solely on the dimensionless parameter $\zeta$.
From this expansion, one can obtain an analytic solution for $S_{\phi}$. 
To leading order in $(E_{\rm gap}-\epsilon)/|\Delta|$ one finds
\begin{equation} 
S_{\phi}=\frac{4}{15}\frac{\alpha^{5/2}}{\beta^2}\left(\frac{
E_{\rm gap}-\epsilon}{|\Delta|}\right)^{5/4}.
\label{eq:1Daction}
\end{equation}
This approximation is shown in Fig.~\ref{fig:action} along with the exact 
result obtained by numerical integration. Note that the action vanishes 
exactly at the gap. For completeness we give the explicit form of the bounce
solution
\begin{eqnarray*}
\phi(x)-\phi_{\sc ag}={\alpha \over \beta}{1\over \cosh^2(x/2r_0)}\;,
\end{eqnarray*}
where the extent of the instanton is set by
\begin{equation} \label{eq:r0}
r_0(\epsilon)={\xi\over \alpha^{1/2}}\left(\frac{|\Delta|}{E_{\rm gap}
-\epsilon}\right)^{1/4}.
\end{equation}
Indeed the size of the instanton is easily understood from the quadratic 
``stiffness'' term in Eq.~(\ref{eq:approxV}). Thus one finds that, while
the overall scale is set by the superconducting coherence length $\xi$, 
the size of the droplet diverges both as $\epsilon$ approaches 
$E_{\rm gap}$ and, noting that $\alpha\sim(1-\zeta^{2/3})^{1/4}$, as 
one approaches the gapless phase $\zeta\to 1$.

\begin{figure}[hbtp]
\centerline{\epsfxsize=3in\epsfbox{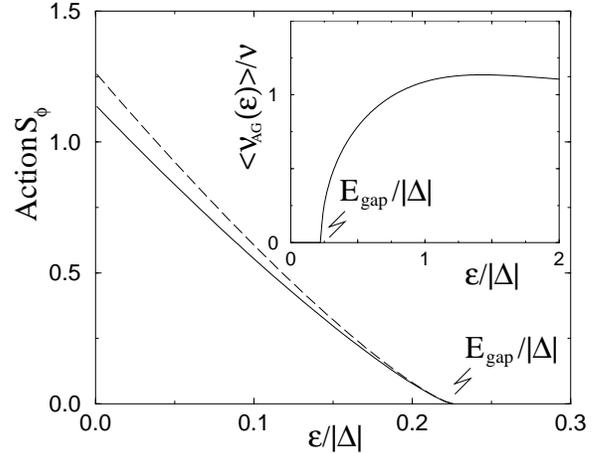}}\bigskip
\caption{Action $S_\phi$ for $\zeta=0.5$ obtained numerically (solid curve) 
together with the expansion in $(E_{\rm gap}-\epsilon)/|\Delta|$ (dotted 
curve) as determined by Eq.~(\ref{eq:approxV}). Note that the action 
vanishes as 
$\epsilon\to E_{\rm gap}$. The AG solution for the DoS is shown inset.}
\label{fig:action}
\end{figure}

This completes the analysis of the saddle-point solution together with the
corresponding action. However, as this level we are presented with two
problems: 
\begin{itemize}

\item the contribution of a second saddle point would seem to spoil 
the normalization condition $\langle{\cal Z}[0]\rangle_{V,S}=1$, which 
should be preserved within the saddle point approximation;

\item confined to the line ${\rm Im}\;\theta=\pi/2$, when substituted
into the DoS source~(\ref{eq:dos_source}), the bounce configuration does 
not appear to generate states!

\end{itemize}
The resolution of both problems lies in the nature of the fluctuations around
the symmetry broken mean-field solution. These field fluctuations can be 
separated into ``radial'' and ``angular'' contributions. The former involve 
fluctuations of the diagonal elements $\hat{\theta}$, while the latter 
describe rotations including those Grassmann transformations which mix the 
{\sc bf} sector~\cite{classD_note}. Both classes of fluctuations play a 
crucial role.

\subsection{Fluctuations} 
\label{sec:fluct}

Before turning to the technical analysis, let us outline qualitatively
the influence of the fluctuations around the mean-field. As usual, 
associated with radial fluctuations around the bounce, there exists a 
zero mode (due to translational invariance of the solution), and a 
negative energy mode. The latter, which necessitates a $\pi/2$ rotation 
of the corresponding integration contour to follow the line of steepest 
descent (c.f. Ref.~\cite{coleman} and see below), has two effects: firstly 
it ensures that the non-perturbative contributions to the local DoS are 
non-vanishing, and secondly, that they are positive. Turning to the angular 
fluctuations, the breaking of supersymmetry is accompanied by the 
appearance of a Grassmann zero mode separated by a gap from higher 
excitations which restores the global supersymmetry (c.f. spin symmetry 
breaking in a ferromagnet of finite extent). The zero mode ensures that 
the symmetry broken inhomogeneous saddle-point configurations respect 
the normalization condition $\langle{\cal Z}[0]\rangle_{V,S}=1$. 

To formally investigate the fluctuation determinant, let us implement the 
rational parameterization 
\begin{equation} 
Q=R\sigma^{\sc ph}_3\otimes\sigma^{\sc cc}_3\frac{1+iP}{1-iP}R^{-1},
\label{eq:ratparam}
\end{equation}
where the condition $P=\sigma^{\sc ph}_2\Gamma P^T \Gamma^{-1} 
\sigma^{\sc ph}_2$ with
\begin{eqnarray*}
\Gamma\equiv \sigma_3^{\sc cc}\gamma =E_{\sc bb}\otimes\sigma^{\sc cc}_1 - 
iE_{\sc ff}\otimes\sigma^{\sc cc}_2,
\end{eqnarray*}
is imposed by (\ref{eq:Qsym}). With this choice, a variant of that 
used in Ref.~\cite{fe}, the measure is trivial~\cite{efetov}. 
$R$ rotates $Q$ from the metallic saddle point $\sigma^{\sc ph}_3\otimes
\sigma^{\sc cc}_3$ to the bounce configuration:
\begin{eqnarray*}
R(x)=\exp\left[\frac{1}{2}\sigma^{\sc ph}_2\otimes\sigma^{\sc cc}_3\;
\hat\theta(x)\right].
\end{eqnarray*}
Defining
\begin{eqnarray*}
P = \pmatrix{C & A \cr B & \Gamma C^T \Gamma^{-1}}_{\sc ph}, 
\end{eqnarray*}
where $A=-\Gamma A^T \Gamma^{-1}$ and $B=-\Gamma B^T \Gamma^{-1}$, the 
condition $Q^2=\openone$ requires $[\sigma^{\sc ph}_3\otimes
\sigma^{\sc cc}_3,P]_+=0$, or
\begin{eqnarray*}
[\sigma^{\sc cc}_3,C]_+=[\sigma^{\sc cc}_3,A]_-=[\sigma^{\sc cc}_3,B]_-=0 \;.
\end{eqnarray*}
Thus the matrix field $C$ is off-diagonal in the {\sc cc} space, while 
$A$ and $B$ are diagonal. In fact, the field fluctuations $C$ describe 
the low-energy quantum interference effects --- the soft Class D modes. 
However, the fluctuations contained within the fields $C$ are oblivious 
to the supersymmetry breaking. We will therefore deal only with the 
fluctuations that are parameterized by $A$ and $B$. (Moreover, we will 
neglect the massive spin triplet fluctuations.)

With the explicit parameterization~\cite{foot1}
\begin{eqnarray*}
A&=&\pmatrix{s_{\sc b} & 0 & \bar{\eta}_A & 0 \cr
             0 & -s_{\sc b} & 0 & \eta_A \cr
             -\eta_A & 0 & s_{\sc f} & 0 \cr
             0 & \bar{\eta}_A & 0 & -s_{\sc f} } \\     
B&=&\pmatrix{-s_{\sc b}^* & 0 & \bar{\eta}_B & 0 \cr
             0 & s_{\sc b}^* & 0 & \eta_B \cr
             -\eta_B & 0 & s_{\sc f}^* & 0 \cr
             0 & \bar{\eta}_B & 0 & -s_{\sc f}^* }\;,
\end{eqnarray*}
an expansion of the action (\ref{eq:sigmod}) to quadratic order around the 
bounce configuration
\begin{eqnarray*}
\hat\theta(x)=\pmatrix{\theta_1(x) & 0 \cr 0 & \theta_{\sc ag}}_{\sc bf}.
\end{eqnarray*}
obtains 
\begin{eqnarray*}
S=4\pi\nu L_W (D|\Delta|)^{1/2} (S_\phi+S_{\rm q})
\end{eqnarray*}
%
%
where $S_\phi$ represents the contribution from the saddle-point 
alone (\ref{eq:realact}), and 
\breakon
\begin{eqnarray} 
S_{\rm q} &=& \int du \left[(|\partial 
s_{\sc b}|^2 +|\partial s_{\sc f}|^2) + V^\prime(\phi) {s^\prime_{\sc b}}^2 
+ V^{\prime\prime}(\phi) {s^{\prime\prime}_{\sc b}}^2 + 
V^\prime(\phi_{\sc ag}) {s^\prime_{\sc f}}^2 + V^{\prime\prime}(\phi_{\sc ag}) 
{s^{\prime\prime}_{\sc f}}^2\right] \nonumber \\
& & \qquad \qquad \qquad +\int 
du \left[(\partial \bar\xi_+\partial \xi_+ +\partial \bar\xi_-\partial \xi_-) 
+ V_+\bar\xi_+\xi_+ + V_-\bar\xi_-\xi_- \right],
\label{eq:quadaction}
\end{eqnarray}
\breakoff
\noindent
where $s_{\sc b/f}\equiv s^\prime_{\sc b/f}+is^{\prime\prime}_{\sc b/f}$, 
$\xi_\pm\equiv(\eta_B\pm\eta_A)/\sqrt{2}$, $\bar{\xi}_\pm\equiv(\bar{\eta}_B\pm
\bar{\eta}_A)/\sqrt{2}$, $u\equiv x/\xi$ and $\theta_{\sc ag}=i\pi/2+
\phi_{\sc ag}$. Here the various potentials are given by
\begin{eqnarray*}
V^\prime(\phi) &=& 2\left(\cosh\phi-{\epsilon\over |\Delta|}\sinh\phi-\zeta
\cosh 2\phi\right)\\
V^{\prime\prime}(\phi) &=& (\partial\phi)^2 + 2\left(\cosh\phi-{\epsilon
\over |\Delta|}\sinh\phi-\zeta\sinh^2\phi\right),
\end{eqnarray*}
and
\begin{eqnarray*}
&&V_{\pm}(\phi)={(\partial\phi)^2\over 4} +\cosh\phi+
\cosh\phi_{\sc ag}\nonumber \\
&& \quad -{\epsilon\over |\Delta|}(\sinh\phi+\sinh\phi_{\sc ag})
-{\zeta\over 2}\left(\cosh 2\phi + \cosh 2\phi_{\sc ag}\right)\nonumber\\ 
&& \quad \quad -\zeta\left(\sinh\phi\sinh\phi_{\rm AG} \mp 
\cosh\phi\cosh\phi_{\rm AG}\right).
\end{eqnarray*}

As a check, let us consider the conventional saddle point. At $\epsilon=0$ 
(for simplicity), 
$\phi=\phi_{\sc ag}=0$, and the quadratic action assumes the form
\begin{eqnarray*} 
&&S_{\rm q}=\int du \left(|\partial s_{\sc b}|^2 +|\partial s_{\sc f}|^2 
+ \partial\bar\xi_+\partial\xi_++\partial\bar\xi_-\partial\xi_-\right) 
\nonumber \\
&& \qquad \qquad +2\int du\Big[|s_{\sc b}|^2+|s_{\sc f}|^2+\bar\xi_+\xi_+
+\bar\xi_-\xi_-\\ 
&& \qquad \qquad \qquad -\zeta({s^\prime_{\sc b}}^2+{s^\prime_{\sc f}}^2 
 +\bar\xi_-\xi_-) \Big].
\end{eqnarray*}
This action is manifestly supersymmetric and, therefore, performing the 
integrations over all fields gives unity. (Moreover, since $\zeta<1$ 
in the gapped phase, the integral is manifestly convergent.) This in why 
the usual AG DoS is just given by evaluating the source 
(\ref{eq:dos_source}) at the supersymmetric AG saddle point.

Since $V^\prime(\phi_{\sc ag})$ and 
$V^{\prime\prime}(\phi_{\sc ag})$ are both positive definite, 
integration over the Fermion-Fermion degrees of freedom merely generates 
some (weakly energy dependent) positive prefactor. The Boson-Boson sector 
is more interesting. In particular, it is straightforward to verify that 
the action for the components $s^\prime_{\sc b}$ simply reflects the 
longitudinal variation 
\begin{eqnarray*}
\frac{1}{2}\int dx\int dx' \Delta\theta_1(x) 
\frac{\delta^2 S}{\delta\theta_1(x)\delta\theta_1(x')} \Delta\theta_1(x')
\end{eqnarray*}
with $2s^\prime_{\sc b}=\Delta\theta_1$. Thus the action for 
$s^\prime_{\sc b}$ is the one that could have been written down from 
the outset: it is that of the fluctuations of $\theta_1$ around the bounce 
instanton discussed in section~\ref{sec:inst}. To address the influence
of this class of fluctuations we can draw on the standard 
literature~\cite{coleman}.

To perform the Gaussian integration over the fields $s^\prime_{\sc b}$, we
form the expansion
\begin{eqnarray*}
s^\prime_{\sc b}(x)=\sum_{\rm n} a_{\rm n} \varphi_{\rm n}(x)
\end{eqnarray*}
in terms of the eigenfunctions $\varphi_{\rm n}$ of the quadratic operator in 
the action for $s^\prime_{\sc b}$. Now the action for $s^\prime_{\sc b}$
exhibits a zero mode $\varphi_1\sim\partial\theta_1$ due to translational 
invariance of the action (the bounce can be positioned anywhere in space). 
This zero mode must be accommodated by the introduction of a collective 
coordinate which in turn introduces a Jacobian factor associated with 
the change of variables from $a_1$ to the collective coordinate. 
Furthermore, since the instanton is a bounce, the zero mode $\varphi_1$ 
has a node, and hence there exists a ground state of $V^\prime$ 
with negative energy. This requires the contour for the integration 
variable $a_0$ to be deformed away from the real axis --- it takes a right 
turn at zero and heads in the negative imaginary direction. In the context 
of the contour drawn in Fig.~\ref{fig:bounce_contour}, this is because we 
have to get back to the real axis for $\theta_1$.

This deformation of the integration contour has a profound consequence. 
While the contribution to the sub-gap states from the bounce solution
alone vanishes (recall that the instanton was confined to the line 
${\rm Im}\;\theta_1=\pi/2$, so that $\cosh\theta_1$ remains imaginary), 
the rotation of the integration contour introduces a factor of $i$ 
resulting in an imaginary contribution to the Green function below the gap.
This is turn would signify a non-zero sub-gap DoS. The mechanism operates
in the related context of Landau levels broadened by disorder and discussed 
by Efetov and Marikhin~\cite{em} (see also an earlier paper by 
Affleck~\cite{affleck}).

It is reassuring to note that the deformation of the integration contour,
which is constrained to return to the undeformed Bosonic contour, is 
unambiguous, and gives rise to only positive definite contributions to 
the DoS. Finally, as usual, since the contour only runs over half a 
Gaussian for $a_{\rm 0}$, its contribution yields a factor of $1/2$.

\begin{figure}[hbtp]
\centerline{\epsfxsize=3in\epsfbox{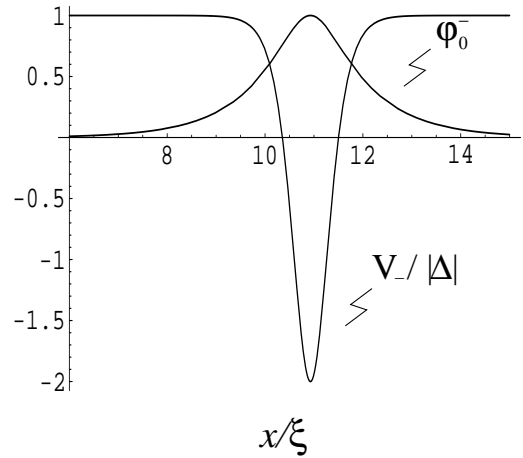}}
\caption{Spatial dependence of the Grassmann zero mode $\varphi^-_0$ for
$\epsilon=0$ and $\zeta=0.5$ together with the potential $V_-$ (scaled by 
$|\Delta|$) that binds it. 
}
\label{fig:goldstino}
\end{figure}

This completes the discussion of the contribution to the functional 
integral from the Bosonic degrees of freedom. Finally we turn to the 
role of the Grassmann fluctuations. Considering the supersymmetric 
structure of the action, it is immediately clear that supersymmetry
breaking must be accompanied by the existence of zero mode in the 
Grassmann sector which restores the supersymmetry. Indeed, an inspection
of the potential $V_-$ identifies a zero energy eigenfunction gapped 
from the others (see Fig.~\ref{fig:goldstino})~\cite{foot2}.

This situation may be compared with the analysis of Andreev and 
Altshuler~\cite{aa}, who identified a stationary phase saddle-point 
that determines the oscillatory part of energy level correlation 
functions for normal diffusive conductors. Again this 
saddle-point is more correctly a manifold of points related by 
supersymmetry, but the fact that it is spatially uniform guarantees that 
the spectrum of Grassmann fluctuations is truly gapless. 

This zero mode indicates the existence of a degenerate manifold 
parameterized by a Grassmann coordinate. The integration of unity over 
this Grassmann coordinate yields zero, which ensures the necessary 
normalization condition $\langle {\cal Z}[0]\rangle_{V,S}=1$ is met. 
A non-zero result can be obtained only if there is a prefactor (or source 
term) that breaks supersymmetry. Making use of the DoS source 
(\ref{eq:dos_source}), we expand in $P$ to find the lowest order part 
quadratic in the Grassmanns (at order $P^2$)
\begin{eqnarray*}
&&{\rm str}\left(\sigma_3^{\sc bf}\otimes\sigma_3^{\sc ph}\otimes
\sigma_3^{\sc cc}Q(x)\right)\simeq \\ 
&& 8i(\sinh\phi(x)-\sinh\phi_{\sc ag})\times \left(\bar\xi_+(x)\xi_+(x)+
\bar\xi_-(x)\xi_-(x)\right).
\end{eqnarray*}
From this result, we see that the local sub-gap DoS remains non-zero 
only in the vicinity of the instanton where the supersymmetry is broken. 

Thus, taking into account Gaussian fluctuations and zero modes, one obtains 
the non-perturbative, one instanton contribution to the sub-gap DoS:
\breakon
\begin{eqnarray} 
\label{eq:dosresult}
{\langle\nu(\epsilon)\rangle_{V,S}\over 4\nu}\sim\; (-i|K|)\ \int dx\;
i(\sinh\phi(x)-\sinh\phi_{\sc ag})\ |\varphi_0^-(x)|^2
\ \sqrt{LS_\phi\over\xi}\; \exp\left[-4\pi\nu L_{\rm w}
\sqrt{D|\Delta|} S_\phi\right]\;,
\end{eqnarray}
\breakoff
\noindent
where the factor $\sqrt{L S_\phi/\xi}$ represents the Jacobian associated 
with the introduction of the collective coordinate~\cite{coleman}, $-i|K|$ 
is the overall factor arising from the non-zero modes, and the Grassmann 
zero mode wavefunction $\varphi^-_0$ is normalized such that
\begin{eqnarray*}
\int dx |\varphi^-_0|^2 = 1.
\end{eqnarray*}
Here we have assumed that the $s^{\prime\prime}_{\sc b}$ integration only 
contributes to the positive prefactor. We have checked this numerically for 
a few cases and expect a general statement could be made by moving to a 
different parameterization.

Eq.~(\ref{eq:dosresult}) is the main result of this section. Note the 
non-perturbative nature of the result, both in the coupling constant 
$g^{-1}$ of the $\sigma$-model, and the (dimensionless) spin scattering 
rate $\zeta$. 

\subsection{Sub-gap States in Dimensions $d>1$} 
\label{sec:hidim}

The calculation above was tailored to the consideration of the quasi 
one-dimensional geometry. The generalization to higher dimensions follows
straightforwardly. In particular, it is necessary to seek inhomogeneous
solutions of the saddle-point equation~(\ref{eq:firstint}) where the 
gradient operator must be interpreted as the higher dimensional 
generalization. Generally, this equation must be solved numerically. However,
for energies $\epsilon$ in the vicinity of the gap $E_{\rm gap}$, an
analytic expression for the energy scaling can be obtained.

Using the approximation to $V_{\sc r}[\phi]$~(\ref{eq:approxV}) valid when
$(E_{\rm gap}-\epsilon)/|\Delta|\ll 1$, the exponential dependence of the
sub-gap DoS can be deduced in higher dimension. In this limit, dimensional 
analysis of the cubic equation of motion yields the scaling form
\begin{eqnarray*}
\phi({\bf r})-\phi_{\sc ag}(\epsilon)=\frac{\alpha}{\beta}f({\bf r}/r_0)\;,
\end{eqnarray*}
where $r_0$ is the characteristic length defined by Eq.~(\ref{eq:r0}).
When substituted back into the action, one finds that the DoS depends
exponentially on the parameter $4\pi g(\xi/L)^{d-2} S_\phi$ where
\begin{eqnarray} {\label{eq:hidim}}
S_\phi=a_d\ \zeta^{-2/3} (1-\zeta^{2/3})^{-(2+d)/8}\left({E_{\rm gap}-
\epsilon\over |\Delta|}\right)^{(6-d)/4}.
\label{eq:genres}
\end{eqnarray}
Here $g=\nu D L^{d-2}$ denotes the bare dimensionless conductance of the
normal system, and $a_d$ is a numerical constant ($a_1=8\ {}^4\sqrt{24}/5$) 
In particular, the exponent depends linearly on the energy separation from 
the gap in two dimensions.

\subsection{Numerics} 
\label{sec:numerics}

To assess the validity of the approximations used in obtaining the
results above we have investigated numerically the DoS in the vicinity
of the mean-field energy gap using a numerical diagonalisation of a
non-interacting tight-binding Bogoluibov Hamiltonian with disorder in
both the on-site potential matrix elements and in the spin impurity
scattering potential. Taking a 
two-dimensional lattice of size $22\times 22$ site with an on-site
disorder taken from the range of $\epsilon\in [-3,3]$ measured in
units of the hopping matrix element, $|\Delta|=1$ measured in the same
units, and various strengths of the spin impurity potential, the DoS
is shown in Fig.~\ref{fig:2d_tail1}. Notice that, as the strength of
the spin impurity potential is increased, the quasi-particle energy
gap is quenched. 

\begin{figure}[hbtp]
\centerline{\epsfxsize=3in\epsfbox{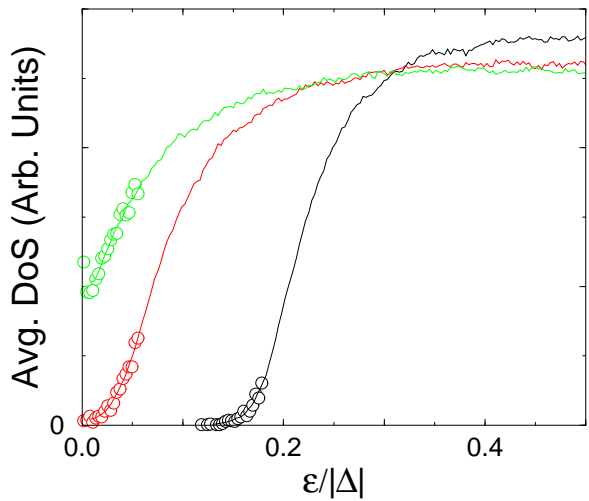}}\bigskip
\caption{Variation of the DoS for a weakly disordered two-dimensional
  tight-binding Bogoluibov
  Hamiltonian in the presence of magnetic impurities. The data is
  shown for increasing strengths of the magnetic impurity
  potential. At the lowest energies the data points are shown as open
  circles to emphasize the fine structure. Notice that in the phase
  where there is predicted to be a gap in the mean-field theory the
  DoS shows a tail extending in the sub-gap region. When the
  quasi-particle energy gap is fully suppressed, the DoS shows an
  upturn at very low energies which is compatible with the
  renormalisation due to quantum interference processes predicted by
  the class D theory.}
\label{fig:2d_tail1}
\end{figure}

For the weakest magnetic impurity potential, we have expanded the
region in the vicinity of the mean-field energy
gap. Fig.~\ref{fig:2d_tail2} shows a exponential scaling of the
`sub-gap' DoS with an exponent which depends linearly on energy as
predicted by the theory above.

\begin{figure}[hbtp]
\centerline{\epsfxsize=3in\epsfbox{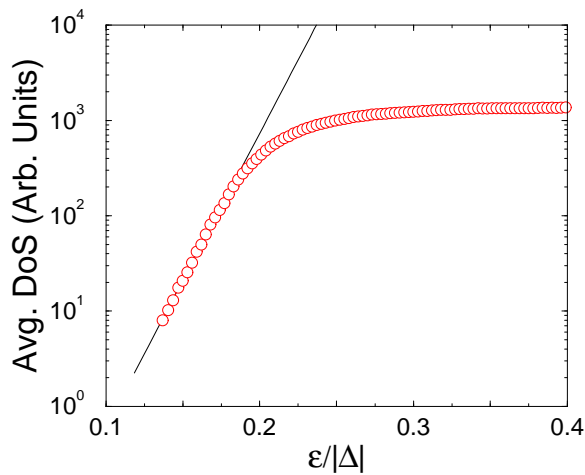}}\bigskip
\caption{Variation of the DoS in the vicinity of the mean-field energy
  gap. The data is shown as open circles and is compared to an
  exponential fit of the data shown as a solid curve. Notice that the
  DoS depends exponentially on the energy difference as predicted by
  the two-dimensional version of the theory.}
\label{fig:2d_tail2}
\end{figure}

\section{Zero dimensional problems and universality}
\label{sec:zerod}

In the previous section we considered the instanton contribution to the 
sub-gap DoS in the infinite system. For completeness let us now consider 
the zero dimensional case that obtains when $r_0$, the size of the 
instanton, exceeds the system size $L$, which will happen when $\epsilon$ 
approaches close enough to $E_{\rm gap}$ from below in any finite
system. In this limit one can clearly not fit an instanton inside the system. 
Leaving aside the practical relevance of this situation, theoretical 
motivation is provided by a recent paper~\cite{bna} that explored this 
regime using a random matrix analysis.

However, before turning to the consideration of the zero-dimensional
limit of the present problem let us first try to draw some intuition
from a closely related investigation of a different system comprised 
of a normal quantum dot contacted to a superconducting terminal (as 
shown in Fig.~\ref{fig:dot}). In such a geometry it is well
established (see e.g. Ref.~\cite{mbfb}) that the proximity effect 
induces a gap in the DoS of the normal dot. Indeed, in
Ref.~\cite{mbfb}, the integrity of the gap is proposed as a signature 
of irregular or chaotic dynamics inside the dot. Now near the gap edge 
the DoS of the dot takes the singular form
\begin{equation} \label{eq:gendos}
\nu(\epsilon>E_{\rm gap})\simeq\frac{1}{\pi L^d}\sqrt{\frac{\epsilon
-E_{\rm gap}}{\Delta_{\rm g}^3}}\;,
\end{equation}
with $E_{\rm gap}=cN\delta$ and $\Delta_{\rm g}=c'N^{1/3}\delta$,
where $c\approx 0.048$ and $c'\approx 0.068$, $\delta=1/L^d\nu$ 
denotes the single particle level spacing, and $N$ is the number of 
fully transmitting channels in the contact. 

However, the location of the gap edge relies on a mean-field analysis
of the coupled system. In Ref.\cite{vbab} Vavilov {\em et al.} have 
argued that optimal
fluctuations of the impurity potential give rise to gap fluctuations. 
The hypothesis introduced in Ref.~\cite{vbab} is that the spectral 
statistics near a gap edge are universal. This allows a random matrix 
theory analysis of gap fluctuations and leads to the following expression
for the sub-gap DoS,
\begin{eqnarray} \label{eq:gentail}
{\nu(\epsilon<E_{\rm gap})\over \nu}\sim \exp\left[-\frac{2}{3}
\left(\frac{E_{\rm gap}-\epsilon}{\Delta_{\rm g}}\right)^{3/2}\right].
\end{eqnarray}

Now the AG mean-field solution for a superconducting quantum dot with
magnetic impurities also predicts the existence of a square root edge
(see Eq.~(\ref{nuexp}) and inset of Fig.~\ref{fig:action}). 
Then, when recast in the form of 
Eq.~(\ref{eq:gendos}), it is pertinent to ask whether the expression for
the sub-gap DoS coincides with Eq.~(\ref{eq:gentail}) in the zero 
dimensional limit. This is the situation addressed in~Ref.\cite{bna}. 

In the following we will show that the universal
result~(\ref{eq:gentail}) is explicitly recovered by the present
theory. Moreover, in doing so, we will expose the origin of the
universal structure reported in Ref.~\cite{vbab} and describe its 
implications for universality of the $d>0$ problem.

\begin{figure}[hbtp]
\centerline{\epsfxsize=3in\epsfbox{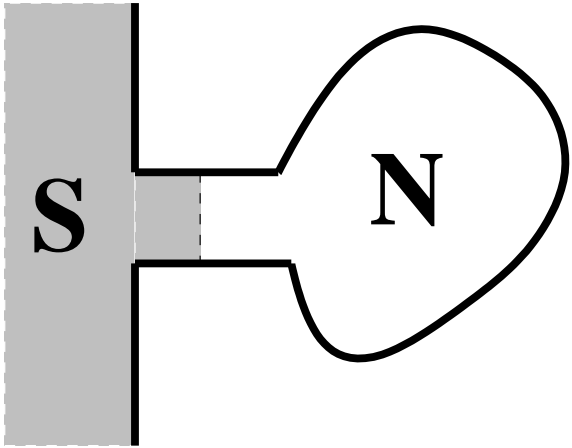}}\bigskip
\caption{Metallic quantum dot coupled to superconducting leads}
\label{fig:dot}
\end{figure}

\subsection{Superconducting dot with magnetic impurities}

Let us therefore consider explicitly the action of a superconducting
grain in the presence of a weak magnetic impurity potential. When
$r_0\gg L$ only the zero spatial mode contributes significantly to the 
action (\ref{eq:sigmod}). In this limit, the action assumes the zero 
dimensional form
\begin{eqnarray} 
&&S[Q]=\frac{i\pi}{2\delta} {\rm str}\left[\left(\epsilon_-
\sigma^{\sc cc}_3+|\Delta|\sigma^{\sc ph}_2\right)\sigma^{\sc ph}_3 Q
\right]\nonumber \\
&&\qquad \qquad \qquad +\frac{\pi}{24\tau_s\delta} {\rm str}\left(Q
\sigma^{\sc ph}_3\otimes {\bf \sigma^{\sc sp}}\right)^2,
\label{eq:sigmod0}
\end{eqnarray}
where, as usual, $\delta$ denotes the single-particle level-spacing. As in 
the higher
dimensional problem, a variation of the action with respect to $Q$ obtains 
a mean-field equation, now without spatial variation. Parameterizing
the saddle-point equation as in section~\ref{sec:ag}, we obtain the 
zero-dimensional Usadel or AG mean-field equation~(\ref{eq:usadel})
\begin{eqnarray*}
2i\left(\cosh\hat\theta-\frac{\epsilon}
{|\Delta|}\sinh\hat\theta\right)-\zeta\sinh(2\hat\theta)=0.
\end{eqnarray*}
From this equation, we can identify the usual AG solution which in turn
recovers the AG phenomenology. 

The inclusion of bounce configurations in the previous calculations was 
based upon the observation that, although the contribution they make is 
exponentially small, they are the least action configurations on the part 
of the contour that gives a finite sub-gap DoS. In the zero-dimensional case 
we are spared having to think about the problem in function space. The 
action is proportional to the potential of Fig.~\ref{fig:bouncepot}. 
The correct contour thus passes through the maximum of the potential 
(minimum of the action) corresponding to the usual AG saddle point, and 
turns away from the real $\phi$ axes (i.e. the line ${\rm Im}\;\theta_1=
\pi/2$) at the minimum of the potential 
\begin{eqnarray*}
\phi_{\sc 0d}(\epsilon)=\phi_{\sc ag}(\epsilon)+{2\alpha\over 3\beta}
\sqrt{E_{\rm gap}-\epsilon\over |\Delta|},
\end{eqnarray*}
(marked in 
Fig.~\ref{fig:bouncepot}). This part of the contour, parallel to the 
imaginary axes, gives a contribution to the DoS, and the second saddle 
point is in fact a \emph{maximum} on this portion by analyticity.
Following the same arguments as in section~\ref{sec:inst}, this 
solution is inaccessible to the Fermionic
contour. We find that the sub-gap DoS near $E_{\rm gap}$ in the 
zero-dimensional case scales as
\begin{eqnarray*}
{\nu(\epsilon<E_{\rm gap})\over\nu}\sim \exp\left[- 
{4\pi\over 27}{|\Delta|\over\delta} {\alpha^3 \over \beta^2}\left({E_{\rm gap}-
\epsilon\over |\Delta|}\right)^{3/2}\right]\;,
\end{eqnarray*}
where $\alpha$ and $\beta$ are the coefficients defined in 
Eq.~(\ref{eq:coeff}). We note that the general result for the energy 
dependence of the exponent written down in section \ref{sec:hidim} 
for dimensions $d\geq 1$ applies also for $d=0$. 

To establish contact 
with the universal result given in Eq.~(\ref{eq:gendos}) it is helpful to 
recast the result in a modified form. To do this we note that, in the 
vicinity of the mean-field gap edge, the DoS can be expanded as~\cite{Park}
\begin{eqnarray*}
&&{\nu_{\sc ag}(\epsilon>E_{\rm gap})\over 4\nu}\\
&&\qquad \qquad \simeq\sqrt{2\over 3}
\;\zeta^{-2/3}(1-\zeta^{2/3})^{-1/4}\left(\frac{\epsilon-E_{\rm gap}}
{|\Delta|}\right)^{1/2}\;.
\end{eqnarray*}
Then, if we define 
\begin{eqnarray*}
\Delta_{\rm g}^{-3/2}\equiv{4\pi\over\delta}\sqrt{\frac{2}{3|\Delta|}}
\zeta^{-2/3}(1-\zeta^{2/3})^{-1/4}\;,
\end{eqnarray*}
the mean-field DoS can be brought to the form of Eq.~(\ref{eq:gendos}), 
and the sub-gap DoS takes the universal form~\cite{foot3}
\begin{eqnarray} \label{eq:gentail2}
{\nu(\epsilon<E_{\rm gap})\over \nu}\sim \exp\left[-\frac{4}{3}\left(
\frac{E_{\rm gap}-\epsilon}{\Delta_{\rm g}}\right)^{3/2}\right]\;.
\end{eqnarray}
Leaving aside a spurious (yet systematic --- see below) factor of 
$2$~\cite{factor2}, the
sub-gap DoS obtained above coincides with the universal expression shown in
Eq.~(\ref{eq:gentail}).

Once again, to assess the validity of the approximation scheme that
leads to the universal result above, we have looked for the scaling in
a numerical investigation of a random matrix Hamiltonian with the same
symmetry. In the sub-gap region, Fig.~\ref{fig:rm_tail} shows a good 
fit of the DoS to the predicted exponential scaling. 

The rescaling of the DoS above and the appearance of the universal
form suggests that we should revisit the $d$-dimensional result and 
look for a similar rescaling. From Eq.~(\ref{eq:hidim}) it is
straightforward to verify that in this case
\begin{eqnarray*}
{\nu(\epsilon<E_{\rm gap})\over \nu}\sim \exp\left[-\widetilde{a}_d
\left({r_0(\epsilon)\over L}\right)^d\left(\frac{E_{\rm gap}-\epsilon}
{\Delta_{\rm g}}\right)^{3/2}\right]\;\;,
\end{eqnarray*}
where $\widetilde{a}_d$ represents some numerical constant, and $r_0$ 
is the characteristic length defined by Eq.~(\ref{eq:r0}). Finally, by 
defining 
\begin{eqnarray*}
\widetilde{\Delta}_{\rm g}^{-3/2}(\epsilon)\equiv {\delta\over
\widetilde{\delta}(\epsilon)} \Delta_g^{-3/2},
\end{eqnarray*}
where $\widetilde{\delta}(\epsilon)=1/(\nu r_0^d(\epsilon))$ is the level 
spacing inside a region of size $r_0$, the volume dependent prefactor can 
be absorbed into the expression and DoS can be brought to the form
\begin{eqnarray*}
{\nu(\epsilon<E_{\rm gap})\over \nu}\sim \exp\left[-\widetilde{a}_d\left(
\frac{E_{\rm gap}-\epsilon}{\widetilde{\Delta}_{\rm g}}\right)^{3/2}
\right]\;\;,
\end{eqnarray*}
revealing a simple relation between the $d=0$ and $d>0$ problems.

The coincidence of Eqs.~(\ref{eq:gentail}) and (\ref{eq:gentail2}) indeed 
suggests that gap fluctuations are universal. To expose the origin of the
universal scaling within the present formalism, let us consider the quantum
dot geometry of Fig.~\ref{fig:dot} within the $\sigma$-model scheme.

\begin{figure}[hbtp]
\centerline{\epsfxsize=3in\epsfbox{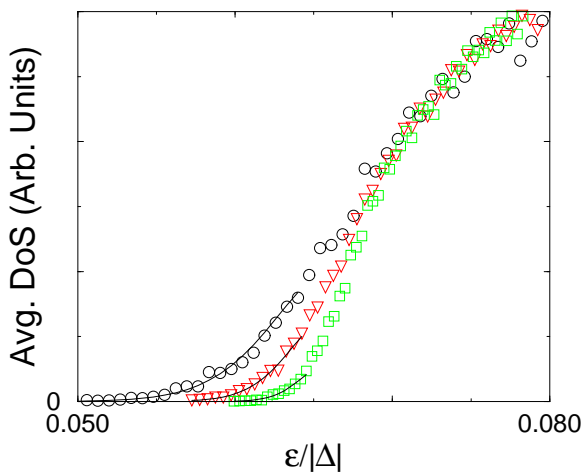}}\bigskip
\caption{DoS for a random matrix Bogoluibov Hamiltonian with a magnetic
  impurity component. Data is shown for three different ensembles
  where only the rank of the matrix is varied ($M=50$, $100$ and $200$). The
  DoS has been rescaled by the average level spacing so that, above
  the mean-field gap edge, the data collapse. As predicted, with this
  rescaling, the width of the tail region decreases with increasing
  size of the random matrix. The data set for $M=200$ is fitted to
  an exponential with a $3/2$ power. The corresponding fit is then
  made with no adjustable parameters to the other two data sets.}
\label{fig:rm_tail}
\end{figure}

\subsection{Quantum dot contacted to a superconductor}

Following Ref.~\cite{vbab}, let us consider a normal quantum dot contacted
to a bulk superconducting terminal with order parameter
$\Delta$. Taking $N$ fully open channels between the dot and the lead,
it is straightforward to show that, in the zero-dimensional limit, the
effective action of the hybrid SN system takes the form
\begin{eqnarray} 
&&S[Q]=-i\frac{\pi\epsilon_+}{2\delta} {\rm str}\left[\sigma^{\sc cc}_3
\otimes\sigma^{\sc ph}_3 Q \right]\nonumber \\
&&\qquad\qquad\qquad\qquad -\frac{N}{2}{\rm str}\left[\ln(1+
Q_{\rm L}Q)\right],
\label{eq:sigmodleads}
\end{eqnarray}
where $Q_{\rm L}$ represents the $Q$-matrix of the superconducting terminal 
(see below), and $Q$ represents the zero-dimensional supermatrix field
within the normal region. The term containing the logarithm describes
the coupling to the lead --- see Ref.~\cite{efetov} for a derivation. 
Even for a diffusive dot, we note that retaining the full logarithm is 
the essential element in the correct treatment of the zero-dimensional 
limit that obtains when $D/L^2\gg N\delta$, a point to which we will 
return later. For a clean superconducting lead, at energies 
$\epsilon$ much less than the bulk superconducting gap $\Delta$, one 
can set $Q_{\rm L}=\sigma_1^{\sc ph}$. 

As usual, to obtain the mean-field expression for the DoS it is necessary to 
minimize the action with respect to variations in $Q$. Doing so, one obtains
the saddle-point equation
\begin{eqnarray*}
i{\pi\epsilon_+\over 2\delta}[Q,\sigma^{\sc cc}_3\otimes\sigma^{\sc ph}_3]
+{N\over 2}[Q,(1+Q_L Q)^{-1} Q_L]=0
\end{eqnarray*}
As usual, applying the {\em Ansatz} that the saddle-point solution is contained
within the diagonal parameterization,
\begin{eqnarray*}
Q=\sigma_3^{\sc cc}\otimes \sigma_3^{\sc ph}\cosh\hat\theta+i
\sigma_1^{\sc ph}\sinh\hat{\theta}, 
\end{eqnarray*}
the saddle-point equation takes the form
\begin{equation} \label{eq:dotsp}
{\pi\epsilon \over \delta}\sinh\hat\theta+{N\over 2}{\cosh\hat\theta\over
1+i\sinh\hat\theta}=0.
\end{equation}
A straightforward analysis of the symmetric saddle-point solution
leads to the mean-field result for the DoS shown in
Eq.~(\ref{eq:gendos}) (remembering that now $\nu(\epsilon)=2\nu\;{\rm Re}
\;\cosh\theta_{\sc mf}(\epsilon)$ as there are no spin degrees of freedom 
in this problem). In particular, one can straightforwardly determine 
$E_{\rm gap}$ by setting $\cosh\theta_{\sc mf}$ to be imaginary. Thus 
$\sinh\theta_{\sc mf}\equiv -ib$ for real $b$ and (\ref{eq:dotsp}) gives
\begin{eqnarray*}
\epsilon(b)={N\delta\over 2\pi}{1\over b}\sqrt{b-1\over b+1}\;.
\end{eqnarray*}
The extremum of this function gives the largest energy corresponding
to a real value of $b$. This occurs at $b=(1+\sqrt{5})/2=1+\gamma$, 
where $\gamma$ is the golden mean, and yields $E_{\rm gap}=
(N\delta/2\pi)\gamma^{5/2}\approx 0.048 N\delta$ as required.

As with the case of magnetic impurities, to explore the influence of gap
fluctuations, it is necessary to seek the symmetry broken saddle-point 
configuration. However, for reasons outlined below, it is possible to 
identify a pattern which implies the universality of the resulting 
analysis:

\subsection{Universalities}

At first glance the situations considered in the two proceeding sections 
are rather different. The actions (\ref{eq:sigmod0}) and 
(\ref{eq:sigmodleads}) would seem not to have much in common. However, a 
simple and general argument may be established to reveal the universal 
character. As before, defining $\theta_{\sc mf}(\epsilon)=i\pi/2+
\phi_{\sc mf}(\epsilon)$, the mean-field DoS for the SN device is 
given by $\nu_{\sc mf}(\epsilon)=2\nu\;{\rm Im}\;\sinh
\phi_{\sc mf}(\epsilon)$, where $\phi_{\sc mf}$ is determined 
by the condition $\delta S/\delta\phi[\phi=\phi_{\sc mf}]=0$. Since 
the DoS displays a square root singularity described by 
Eq.~(\ref{eq:gendos}), the (saddle-point) action near the edge is 
constrained to be of the form
\begin{eqnarray*}
S[\hat{\phi}]=-k\ {\rm str}\left[{1\over 3}\hat{s}^3+\left({\delta\over
2\pi}\right)^2
\left({\epsilon_+-E_{\rm gap}\over\Delta_{\rm g}^3}\right)\hat{s}\right]\;,
\end{eqnarray*}
where $\hat{s}(\epsilon)=\sinh\hat{\phi}(\epsilon)-
\sinh\hat{\phi}_{\sc mf}(E_{\rm gap})$. Here, the elements
$\hat{\phi}={\rm diag}(\phi_{\sc bb},\phi_{\sc ff})$ and 
$\hat{s}={\rm diag}(s_{\sc bb},s_{\sc ff})$ are diagonal in the superspace.
(As one may check, a variation of the action for $\epsilon>E_{\rm gap}$ 
obtains the symmetric mean-field solution 
\begin{eqnarray*}
s_{\sc bb}=s_{\sc ff}=i{\delta\over 2\pi}\sqrt{\epsilon-E_{\rm gap}
\over \Delta_g^3}
\end{eqnarray*}
which in turn recovers the expression~(\ref{eq:gendos}) for $\nu(\epsilon)$.)
Moreover, since the term containing $\hat{s}$ is linear in the energy, we 
can determine the value of $k$ from the knowledge that $\epsilon$ 
appears in the action as $(2\pi\epsilon_+/\delta)\sinh\hat{\phi}$. (It is 
this term that can more generally contain the Dyson index `$\beta$', 
which therefore appears in the general expression for gap fluctuations
described in Ref.~\cite{vbab}.) In the present case, we thus have 
$k=(2\pi\Delta_{\rm g}/\delta)^3$. 

Now, as discussed in the previous section, when $\epsilon<E_{\rm gap}$ 
there exists two saddle-point solutions at 
\begin{eqnarray*}
s_\pm=\pm {\delta\over 2\pi}\sqrt{E_{\rm gap}-\epsilon\over \Delta_g^3}.
\end{eqnarray*}
As before, one of these solutions ($s_-(\epsilon) \leadsto 
\phi_{\sc mf}(\epsilon)$) is associated with the conventional symmetric 
mean-field solution while the other represents a second saddle-point 
accessible only to the Bosonic contour. Taking this second, symmetry broken
saddle-point into account (i.e. setting $s_{\sc bb}=s_+$ and $s_{\sc ff}=s_-$),
one obtains the saddle-point action
\begin{eqnarray*}
S[\hat{\phi}]={4\over 3}\left(\frac{E_{\rm gap}-\epsilon}
{\Delta_{\rm g}}\right)^{3/2}.
\end{eqnarray*}
It is this symmetry broken saddle-point which controls the sub-gap DoS 
and leads to the universal scaling form proposed in Ref.~\cite{vbab}. 
This generalizes the arguments applied to the superconducting dot 
with magnetic impurities.


\subsection{Discussion}

Following on from this discussion, to conclude this section, let us
make two remarks which bare on the universality of the general
scheme. The first of these remarks concerns the integrity of the
scaling of the sub-gap DoS when different impurity distributions are
taken into account. The second remark concerns the
extension of the ideas above to the consideration of the hybrid SN
system beyond the zero-dimensional regime. 

Firstly, for the superconductor with magnetic impurities, one can
generalize the arguments above to show that the energy scaling of the
sub-gap DoS even in the $d$-dimensional case is insensitive to the 
nature of the random impurity distribution. This is in contrast to 
Lifshitz band tail states in semi-conductors where the energy scaling
depends sensitively on this distribution. To understand this, let us
suppose that the distribution of magnetic impurities $J{\bf S}({\bf r})$ is 
not Gaussian $\delta$-correlated, as we assumed throughout, but obeys some 
arbitrary statistics defined by a probability functional 
$P[J{\bf S}({\bf r})]$. When the ensemble average over $J{\bf S}({\bf r})$ 
is performed one would obtain in the $\Psi$-field action a
contribution of the form
\begin{eqnarray*}
\ln\;\left\langle \exp\left[-i\int d{\bf r}\; \bar{\Psi}J{\bf S}
\cdot{\bf \sigma}^{\sc sp}\Psi\right]\right\rangle_P\equiv C[\bar{\Psi}
{\bf \sigma}^{\sc sp}\Psi(\bf r)]\;,
\end{eqnarray*}
which defines $C[\cdots]$, the generating functional of connected correlators 
of $J{\bf S}({\bf r})$. Though this is in general a very complicated and 
indeed non-local functional of $\bar{\Psi}{\bf\sigma}\Psi(\bf r)$, one can 
in principle find a \emph{local} $Q$-field action by including pairings only 
at coincident points, justified by the assumption $(\ell/\xi)^d\ll 1$ about 
the non-magnetic disorder. The mean-field description of this system then 
follows from the homogeneous solution of the saddle-point equation, an Usadel 
equation like (\ref{eq:usadel}) with some potential. Generally 
this potential will have the same characteristics as the real potential of 
(\ref{eq:realpot}) plotted in Fig.~\ref{fig:bouncepot} on the line 
${\rm Im}\;\theta_1=\pi/2$. The central maximum is due to the $|\Delta|$ 
term; the upturn at large $\phi$ arises from the small pair-breaking part, 
and the asymmetry comes from the $\epsilon$ term. Now, if mean-field 
theory leads to a square-root singularity in the DoS (a circumstance
which can be avoided only by a special tuning of parameters), one can
expect that increasing the energy leads to the maximum merging with 
one of the minima according to 
\begin{eqnarray*} 
V_{\sc r}[\phi]\simeq -\alpha\left(\frac{E_{\rm gap}-\epsilon}{|\Delta|}
\right)^{1/2} (\phi-\phi_{\sc mf})^2+ \beta (\phi-\phi_{\sc mf})^3
\end{eqnarray*}
with $\alpha$ and $\beta$ chosen appropriately. Then the analysis of section 
\ref{sec:inhomo} applies. In particular the scaling of the exponent with 
$((E_{\rm gap}-\epsilon)/ |\Delta|)^{(6-d)/4}$ is expected to be 
\emph{universal and independent of the details of the magnetic impurity
potential}. 

Now let us turn to the generality of the present scheme in describing 
`gap fluctuations' in extended hybrid superconductor/normal systems. The
latter has been discussed in a very recent paper by Ostrovsky 
{\em et al.}~\cite{osf}. In this work, the authors developed an 
instanton approach analogous to that employed in the magnetic 
impurity system here to estimate the profile 
of gap fluctuations in the $d$-dimensional SNS system. Now from the
discussion above, it is possible to expose the relation between these 
two works: in the SNS system, the energy gap induced in the normal 
region due to the proximity effect is determined by the Thouless
energy defined as $E_{\sc T}\sim 1/\tau_{\rm dwell}$, where 
$\tau_{\rm dwell}$ is the time required for electrons in the normal 
region to feel the presence of the superconductor~\cite{mbfb}. The 
Thouless energy is determined by $E_{\sc T}\sim \min\{D/L^2,\Gamma 
N\delta\}$ where $\Gamma$ is the transparency of the contact to the 
superconductor ($\Gamma=1$ in the analysis of the zero-dimensional 
system above). 

In the diffusive limit $D/L^2\ll \Gamma N\delta$, at the mean-field 
level, the position of the quasi-particle energy gap is found by 
solving the Usadel equation with the appropriate boundary 
conditions~\cite{gk}. As a result one obtains a square root singularity 
in the DoS. In this case the mean-field solution is itself inhomogeneous.
The sub-gap correction is found by identifying a second inhomogeneous 
instanton configuration that breaks supersymmetry at the level of 
the action~\cite{osf}. Both solutions merge at the mean-field gap.
Now, following the arguments above, it is simple to see how the 
phenomenology of Ref.~\cite{osf} fits into the same general scheme: in
this case the relevant coordinate of $Q$ interpolates between the  
inhomogeneous mean-field solution and the instanton. The result is 
a sub-gap DoS which assumes the familiar form of 
Eq.~(\ref{eq:gentail2}), with appropriately defined geometry 
dependent parameters $E_{\rm gap}$ and $\Delta_{\rm g}$. Naturally the 
introduction of $d_{\perp}$ transverse dimensions gives the expected
energy dependence of $(E_{\rm gap}-\epsilon)^{(6-d_{\perp})/4}$ in 
the exponent.

In the opposite limit $D/L^2\gg\Gamma N\delta$ (not considered in 
Ref.~\cite{osf}), gradients of $Q$ are 
heavily penalized and the coupling to the lead {\em must} be retained 
in its `logarithmic form', with $Q$ being taken as constant in the 
dot. (Indeed, the logarithm is crucial to reproduce even the mean-field 
expression for the DoS~(\ref{eq:gendos}) with the correct coefficients.) 
This is the true zero-dimensional limit treated above. As we have seen, 
with this action, one recovers the known universal expression for the 
spectrum of gap fluctuations below the mean-field edge.

\section{Conclusions}
\label{sec:discuss}

To conclude, we have developed a quasi-classical field theory describing 
a superconductor in the dirty limit with weak magnetic impurity scattering. 
The Abrikosov-Gor'kov mean-field treatment of this system, showing a 
diminished but hard gap in the DoS, can be straightforwardly recovered as
a homogeneous saddle-point of the effective action. Where zero DoS is 
predicted by the mean-field theory, there exist spatially inhomogeneous 
saddle-point configurations that break supersymmetry at the level of the
action. A careful analysis of fluctuations around these instanton 
configurations demonstrate how supersymmetry is restored by a manifold of 
equivalent configurations parametrized by a Grassmann coordinate, but more 
importantly how the configurations give rise to a finite, though 
exponentially small, DoS. In contrast to band tail states in semi-conductors, 
the quasi-particle nature of the sub-gap states leads to universality of 
their properties. 

Finally, let us remark on the connection of the results presented above to 
related problems in the literature. The resulting expression for the DoS 
(\ref{eq:genres}) was found to be non-perturbative in the $\sigma$-model 
coupling $1/g$, which measures the strength of non-magnetic disorder. We 
note that other non-perturbative results in disordered systems have been 
obtained by related instanton calculations. As well as the investigation 
of tail states in semi-conductors~\cite{cardy}, a supersymmetric field 
theory was developed by Affleck~\cite{affleck} (see also Refs.~\cite{em}) 
to investigate tail states in the lowest Landau level. There it was shown 
that tail states correspond to instanton configurations of the 
{\em $\Psi$-field action} (c.f. Ref.~\cite{cardy}). It is also interesting 
to compare the present scheme with the study of `anomalously localized 
states'~\cite{km} (see also, Ref.~\cite{fe}). There one finds that 
long-time current relaxation in a disordered wire is also associated 
with instanton configurations of the $\sigma$-model action. Finally, 
a Lifshitz argument has been applied on the level of the Usadel equation 
in the study of gap fluctuations due to inhomogeneities of the BCS 
interaction~\cite{lo}.

Although we have focussed largely on the question of tail states below
the mean-field quasi-particle gap of a superconductor with magnetic 
impurities, we expect the instanton approach developed here to be more 
widely applicable. Indeed, in this work we have shown the intimate 
connection between the study of sub-gap states in the magnetic impurity 
problem and gap fluctuations in hybrid superconductor/normal structures. 
Furthermore, the same instanton approach describes gap fluctuations in 
superconductors with a quenched inhomogeneous distribution of the 
BCS coupling constant~\cite{lo}, as well as quasi-two dimensional 
superconducting films subject to strong in-plane magnetic 
fields~\cite{meyer}. In both cases the latter are described by a 
mean-field theory which assumes the Abrikosov-Gor'kov form.

More speculatively, it seems likely that the same general scheme can be
employed to study the influence of optimal fluctuations on the nature of 
bulk transitions. For example, in the present system, the transition to bulk 
superconductivity with reducing magnetic impurity concentration will be 
preempted by the nucleation of superconducting islands or droplets 
within the metallic/insulating phase (c.f. Ref.~\cite{ioffe}). Similarly, 
the Stoner transition to a bulk itinerant ferromagnet in a disordered 
system will be mediated by the formation of a droplet phase in which 
islands become ferromagnetic~\cite{aleiner}. In both cases, we expect 
these `droplet phases' to be associated with inhomogeneous instanton 
configurations of the corresponding low-energy action. 

{\sc Acknowledgements}: We are grateful to Alexander Altland,
Alexander Balatsky, Alex Kamenev, Julia Meyer, and Martin Zirnbauer 
for valuable discussions. We are also 
deeply indebted to Dima Khmel'nitskii for bringing this general subject to our 
attention, and to John Chalker for crucial discussions, particularly those 
made at an early stage of this work. One of us (AL) would like to acknowledge 
the financial support of Trinity College.

\end{multicols}


\begin{thebibliography}{99}
\bibitem{classD_note}
We note that the angular fluctuations include both connected components of 
the target manifold associated with the class D modes~\cite{bcsz2}.
\bibitem{foot1} We have used the reality 
conditions $Q_{\sc bb}=\sigma^{\sc ph}_3\otimes\sigma^{\sc cc}_3 
Q_{\sc bb}^{\dagger} \sigma^{\sc ph}_3\otimes\sigma^{\sc cc}_3$ and 
$Q_{\sc ff}=Q_{\sc ff}^{\dagger}$\cite{zirn_class}
\bibitem{foot2} The zero mode 
shown in Fig.~\ref{fig:goldstino} does not look like it generates a 
spatially uniform rotation --- it is in fact a bound state solution of
the potential $V_-$, and asymptotically approaches zero exponentially. 
This is a subtlety that stems from the parametrization~(\ref{eq:ratparam}). 
Were we instead to consider a fluctuation around the instanton of the form
%
\begin{eqnarray*}
Q=TV\sigma^{\sc ph}_3\otimes \sigma^{\sc cc}_3V^{-1}T^{-1},
\end{eqnarray*}
%
where $V$ is the rotation matrix that appears in Eq.~(\ref{eq:ratparam}), 
and $T=e^W$ where 
%
\begin{eqnarray*}
W=\pmatrix{ 0 & w^{\sc bf} \cr w^{\sc fb} & 0}_{\sc bf},\qquad 
w^{\sc bf}=\pmatrix{\kappa & 0 \cr 0 & \bar\kappa}_{\sc cc},
\end{eqnarray*}
%
the zero mode solution would be manifest: spatially uniform fluctuations
$T$ leaves the action unchanged. Though this viewpoint makes the existence 
of the zero mode clearer, it obscures the fact that it is gapped from the 
rest of the fluctuations. This happens because $T$ becomes ineffective 
far from the instanton as $V\sigma^{\sc ph}_3\otimes\sigma^{\sc cc}_3V^{-1}$ 
becomes supersymmetric. This would reflect in the measure associated with 
the fluctuations parameterized by $T$, and amounts to a "stiffness". Because 
the rational parameterization has trivial measure, this effect is visible at 
the level of the action.
\bibitem{foot3} This 
result contradicts that of Ref.~\cite{bna}, most notably in the power law 
dependence of the exponent on the distance from the gap edge ($3/2$ versus 
$2$ in~\cite{bna}).
\bibitem{factor2} 
We note that the factor of $2$ discrepency can be straightforwardly 
accommodated into a redefinition of $\Delta_g^{-3/2}$, or, equivalently,
a rescaling of the mean-field DoS. We therefore expect that this is the 
source of confusion. For reasons outlined in the text, we believe that the
convention adopted here is the consistent one.
\end{thebibliography}
\end{document}